\documentclass[manuscript,screen]{acmart}
\AtBeginDocument{%
  }

\setcopyright{acmlicensed}
\copyrightyear{2018}
\acmYear{2018}
\acmDOI{XXXXXXX.XXXXXXX}
\acmConference[Conference acronym 'XX]{Make sure to enter the correct
  conference title from your rights confirmation email}{June 03--05,
  2018}{Woodstock, NY}
\acmISBN{978-1-4503-XXXX-X/2018/06}
\usepackage{multirow} 
\usepackage{graphicx}
\usepackage{float}

\begin{document}

\title{Detection and Prevention of Process Disruption Attacks in the Electrical Power Systems using MMS Traffic: An EPIC Case}

\author{Praneeta K Maganti}
\email{praneetamaganti@gmail.com}
\orcid{0000-0002-0926-0840}
\affiliation{%
  \institution{Birla Institute of Technology and Science Pilani, Hyderabad Campus}
  \streetaddress{Dept. of Computer Science \& Information Systems}
  \city{Hyderabad}
  \state{Telangana}
  \country{India}
  \postcode{500078}
}

\author{Daisuke Mashima}
\email{daisuke_mashima@sutd.edu.sg}
\affiliation{%
  \institution{Singapore University of Technology and Design}
  \streetaddress{Information Systems Technology and Design}
  \city{Singapore City}
  \state{}
  \country{Singapore}
}

\author{Rajib Ranjan Maiti}
\email{rajibrm@hyderabad.bits-pilani.ac.in}
\affiliation{%
  \institution{Birla Institute of Technology and Science Pilani, Hyderabad Campus}
  \streetaddress{Dept. of Computer Science \& Information Systems}
  \city{Hyderabad}
  \state{Telangana}
  \country{India}
}
\renewcommand{\shortauthors}{Praneeta et al.}

\begin{abstract}
Smart grids are increasingly exposed to sophisticated cyber threats due to their reliance on interconnected communication networks, as demonstrated by real-world incidents such as the cyberattacks on the Ukrainian power grid. In IEC 61850-based smart substations, the Manufacturing Message Specification (MMS) protocol operates over TCP/IP to facilitate communication between SCADA systems and field devices such as Intelligent Electronic Devices (IEDs) and Programmable Logic Controllers (PLCs). Although MMS enables efficient monitoring and control, it can be exploited by adversaries to generate legitimate-looking packets for reconnaissance, unauthorized state reading, and malicious command injection, thereby disrupting grid operations.
In this work, we propose a fully automated attack detection and prevention framework for IEC 61850-compliant smart substations to counter remote cyberattacks that manipulate process states through compromised PLCs and IEDs. A detailed analysis of the MMS protocol is presented, and critical MMS field–value pairs are extracted during both normal SCADA operation and active attack conditions. The proposed framework is validated using seven datasets comprising benign operational scenarios and multiple attack instances, including IEC61850Bean-based attacks and script-driven attacks leveraging the libiec61850 library. Our approach accurately identifies attack-signature-carrying MMS packets that attempt to disrupt circuit breaker status, specifically targeting the smart home zone IED and PLC of the EPIC testbed. The results demonstrate the effectiveness of the proposed framework in precisely detecting malicious MMS traffic and enhancing the cyber resilience of IEC 61850-based smart grid environments.
\end{abstract}


\keywords{Do, Not, Use, This, Code, Put, the, Correct, Terms, for,
  Your, Paper}

\received{20 February 2007}
\received[revised]{12 March 2009}
\received[accepted]{5 June 2009}

\maketitle

\section{Introduction}
The smart grid is not just an electrical system, it is an electric + communication + IT system. Intelligence comes from continuous monitoring, automation, and bidirectional communication and control. These added layers make the grid more efficient, responsive, and capable of balancing supply and demand~\cite{NIST_SmartGridBeginnerGuide, IEEE_Std_2030_2011, NSGM_SmartGrid}. However, the same communication and IT capabilities that enable this intelligence, introduce new dependencies on networked and software-based components. While increased interconnectivity between SCADA systems and field devices in industrial control systems (ICS) has significantly improved operational efficiency, it has also expanded the cyber attack surface, exposing critical infrastructure to a wide range of cyber threats. The real-time control and monitoring in substations heavily rely on protocols like Manufacturing Message Specification (MMS) defined under the IEC 61850 standard, to achieve automation and enabling seamless communication between the programmable logic controllers (PLCs) and intelligent electronic devices (IEDs) that are integral part of smart power grids.  Despite the risks of cyber incidents, MMS was not originally designed to support  strong cybersecurity mechanisms, leaving it vulnerable to manipulation and exploitation. Since MMS messages directly influence the protection relays, circuit breakers and control actions, any malicious interference, such as false trip commands, or altering measurement data can lead to serious operational disruptions or even large-scale power outages. 

Consequently, securing MMS-based communication has become a critical concern in safeguarding the reliability and resilience of modern power grids. 
While prior research has explored anomaly detection in SCADA and IEC 104 networks, the specific characteristics of MMS traffic for detecting attack signatures that disrupt physical processes, remain underexplored. Existing intrusion detection systems often treat MMS as generic TCP traffic, overlooking protocol-specific indicators that could reveal targeted attacks. 
      
Industrial protocols often lack proper security mechanisms and hence attackers can craft and inject malicious packets into the industrial controllers, like Programmable Logic Controllers (PLCs) or Intelligent Electronic Devices (IEDs), irrespective of whether the malicious payload could harm notoriously. 
Further, such protocols vary significantly based on the actual industrial process being monitored and controlled, e.g., water treatment, water distribution, power generation, power distribution or assembly line management. 

CrashOverride (aka Industroyer) is a sophisticated malware framework used in attacks against industrial control systems, most notably the 2016 Ukrainian power-grid incident that caused large Kyiv outage, combined a phishing campaign with protocol-level commands to directly manipulate field devices. The adversary targeted PLCs that respond to IEC-61850 commands, issuing legitimate-looking protocol instructions to change breaker states and force outages. The operator component relied heavily on existing system tools (for example Windows PowerShell) and on legitimate industrial protocol commands like \texttt{CSWI\$CO\$Pos\$Oper} (value of mms.Itemid field in the MMS-PDU) to move laterally and alter PLC state, employing largely fileless techniques that make detection difficult for traditional intrusion-detection systems~\cite{9212128}.
The attackers demonstrated deep protocol knowledge, e.g., MMS/COTP session handling, getNameList enumeration of VMDs and logical nodes (CSWI), and targeted stVal/ctlModel variables, insights that~\cite{salazar2024tale} authors confirmed using their NEFICS sandbox and detailed traffic/behavioral analysis.

Our approach is capable of detecting these kinds of CrashOverride/ Industroyer style attacks, and to our knowledge is the first study that explicitly delves into ICS protocol semantics to extract attack signatures and provide a proactive, protocol-aware detection capability.

\subsection{Motivation}
The North American Electric Reliability Corporation (NERC) Critical Infrastructure Protection (CIP) standards mandate the implementation of security controls such as firewalls, anti-malware solutions, and intrusion detection mechanisms to safeguard critical cyber assets \cite{NERC_CIP005_7}. Specifically, CIP-005-7 (Table R1, Part 1.5) requires entities to employ methods for detecting known or suspected malicious communications, both inbound and outbound, using tools such as intrusion detection systems or application layer firewalls. Likewise, security features, cyber security requirements and best practices for communication protocols defined by IEC61850 standard are provided by IEC 62351-4 and IEC 62351-6 standards~\cite{IEC_TS_62351_100_4_2023, IEC62351_6_2020}. When implemented by organizations utilizing IEC61850 standrd protocols, these mechanisms mitigate many known threats, but adversaries continue to develop new attack strategies that can bypass conventional defenses. 
For instance, the recently disclosed \emph{CVE-2025-0814} identified an Improper Input Validation vulnerability in a Schneider Electric device, where malicious IEC 61850-MMS packets could cause a denial-of-service in network services while leaving the breaker’s core functionality unaffected~\cite{CVE-2025-0814}.
Attackers and defensive testers alike commonly leverage open-source MMS/IEC 61850 stack, such as libIEC61850~\cite{libIEC61850} and IEC61850bean/OpenIEC61850~\cite{beanit_iec61850bean}, to issue both legitimate and malformed control commands (MMS PDUs) directed at field devices, demonstrating how these implementations can be used to test the robustness of MMS message processing and input validation in devices that control electric and other critical infrastructure operations.
This underscores persistent weaknesses in protocol implementations that can be exploited despite existing security controls.
Consequently, continuous monitoring and adaptive detection mechanisms are essential to maintain system resilience. 

\subsection{Our Work and Contributions}
Our work focuses on enhancing intrusion detection capabilities in smart grid, by identifying attack signatures within the Manufacturing Message Specification (MMS) protocol, an essential communication protocol in IEC 61850-based substations. Our approach aims to improve early threat detection and strengthen the overall cybersecurity posture of modern power systems.
We presents a novel methodology for detecting cyberattacks in MMS network traffic by identifying protocol-level attack signatures. Unlike existing anomaly-based approaches, the proposed method leverages field-level inspection of MMS messages to reveal distinctive patterns of malicious behavior from that of normal operations. Accurate detection of attack signatures in MMS network traffic is essential for ensuring the integrity and availability of power grid operations. Recognizing malicious communication patterns before they propagate into control actions, can significantly reduce the risk of cascading failures. We summarize our contributions as follows:
\begin{itemize}
\item Proposed a fully automated MMS attack detection and prevention pipeline that transforms raw pcap inputs into deployable NIDS rules, which significantly reduce manual analysis and rule-generation time.

\item Extracted key MMS protocol fields mms.domainId, mms.itemId, mms.iec61850.timeaccuracy and mms.data.octet-string from both normal and attack traffic to enable fine-grained behavioral differentiation.

\item Constructed separate whitelists for MMS Read and Write operations, representing legitimate (field–value) combinations observed in benign traffic and forming a reliable baseline for anomaly detection.

\item Developed a deviation-based detection algorithm that compares new MMS pcaps against these whitelists and identified attack signatures, to automatically identify abnormal field values indicative of malicious write operations which may constitute zero day attack.

\item Generated precise attack signatures from the extracted abnormal MMS field values and reconstructed attack paths (origin device, target PLC/IED, operation, affected field component), providing high interpretation and accuracy in detection.

\item Created mechanism to automatically generate NIDS rules from the identified signatures, enabling direct deployment and rapid response to emerging MMS-based attacks.

\item We validated the complete pipeline on seven datasets (eight from the EPIC physical testbed and one virtual EPIC dataset), showing robust detection of IEC61850bean tool attacks and detection of previously unseen process disruption attacks with zero false positive rates.
\end{itemize}

\section{Related Works}
Yang et al. \cite{yang2016multidimensional} proposed a multidimensional IDS for IEC 61850 SCADA networks that inspects MMS, GOOSE, and SMV traffic using four detection dimensions: (i) access-control verification based on MAC/IP addresses and port pairs; (ii) protocol whitelisting for authorized IEC 61850-related traffic; (iii) model-based detection using Substation Configuration Description (SCD) files to define normal station- and process-bus behavior, including allowed MMS services, client limits, file-transfer constraints, and timing rules; and (iv) multi-parameter detection that correlates source identity, service type, and temporal behavior. For MMS, the IDS models association, reporting, setting, file-transfer, and timing services, detecting unauthorized write operations when Write/SetDataValues requests originate from non-SCD-authorized hosts. The system relies on source identity, service type, and timing characteristics rather than parsing internal MMS field semantics.

Hong et al. \cite{hong2014integrated} introduced an integrated anomaly detection system combining network-level analysis of GOOSE and SV traffic with system-level physical measurements. Using a cyber-physical testbed and IEEE-39 bus simulations, the system analyzes PCAP-based packet data to detect replay, modification, injection, and DoS attacks. A specification-based whitelist approach validates GOOSE state/sequence behavior and SV packet-rate thresholds, while system-level checks identify abnormal device responses.   Albarakati et al.~\cite{albarakati2021security} developed an network and system management (NSM) data based monitoring platform for a realistic substation model that integrates power system components, protection devices, controllers, and IEC 61850 communication protocols such as MMS, GOOSE and Sampled Values. A two-stage deep learning framework combining recurrent neural networks with autoencoders and ensemble learning is proposed to detect cyberattacks and anomalies. The approach is evaluated using hardware-in-the-loop simulations on the IEEE 9-bus system

Yang et al. \cite{yang2016intrusion} proposed a SCADA-specific IDS evaluated on a 500 kV cyber–physical testbed using real MMS, GOOSE, and SMV traffic. The IDS employs access-control, protocol-based, anomaly-behavior, and multi-parameter detection driven by SCD models to identify unauthorized connections, protocol misuse, abnormal traffic patterns, DoS behavior, and inconsistencies between GOOSE and MMS signaling. Authors in~\cite{yang2022new} proposed anomaly detection for IEC 61850-based smart substation traffic using deep learning techniques. Their framework characterizes network behavior using multiple traffic features extracted over time windows. Discrete Wavelet Transform (DWT) is employed to derive deep features, followed by an improved Locally Linear Embedding (LLE) algorithm for dimensionality reduction. An LSTM-based autoencoder is then used to learn normal traffic patterns and detect anomalies based on reconstruction error. The approach is evaluated using real smart substation traffic captured from a Chinese government testbed, including benign and attack scenarios targeting GOOSE and MMS protocols, such as flooding, replay, malformed packet, and protocol-specific attacks, demonstrating effective and timely detection performance.

Several works focus on data-driven intrusion detection for GOOSE traffic. De et al. \cite{de2025rnn} applied recurrent neural networks (RNN, LSTM, BiLSTM, GRU, BiGRU) to detect replay, masquerade, and injection attacks using RTDS-based hardware-in-the-loop experiments. Features include GOOSE header fields (stNum, sqNum, timeAllowedToLive) and derived temporal differences. Jay et al. \cite{jay2025intelligent} proposed an Intelligent Intrusion Detection Mechanism combining ensemble unsupervised learning with model-based analysis using electrical network information. The system detects MITM, replay, false data injection, suppression, DoS, and composite attacks using stNum, sqNum, timing, and payload measurements, and is validated on both open-source datasets and a real substation testbed.

Baranwal et al. \cite{baranwal2025domain} addressed MMS-specific zero-day and reconnaissance attacks using domain-aware rule-based and machine learning–based IDS designs. The framework detects malicious MMS domain names and IED scanning behavior using heuristic rules and Word2Vec-based semantic embeddings combined with probabilistic anomaly detection models. A related work \cite{baranwal2025critical} focuses on MMS reconnaissance and control modification attacks by analyzing MMS error traffic and malformed packets, generating dedicated datasets using simulated smart grid environments and extracting MMS-specific features from PCAP traces.

Yadgar et al. \cite{yadgar2025detection} proposed a multi-unit monitoring system for the EPIC testbed, generating synthetic attack scenarios from attack-free data to detect synchronization, economic advantage, PLC code manipulation, DoS, load-sharing, and power interruption attacks using rule-based, model-based, and PCA-based detection units.

Cyber–physical IDS approaches have also been explored. Zhu et al. \cite{zhu2020intrusion} proposed a distributed IDS that detects falsified MMS measurements by enforcing Kirchhoff’s Voltage Law and Ohm’s Law across substations, validated on the IEEE 39-bus system. Wang et al. \cite{wang2020detection} introduced an interval observer–based detection and isolation scheme using power-system state measurements for false data injection attacks, employing adaptive interval residuals to handle external disturbances.

\begin{table}[]
\caption{Related Works}
\label{tab:my-table}
\scalebox{0.7}{
\begin{tabular}{|l|l|l|l|l|l|l|l|l|l|l|}
\hline
\begin{tabular}[c]{@{}l@{}}Sr. \\ No\end{tabular} & Ref                                              & Testbed/ Dataset                                                                                                                                                                                                                                            & Data Type                                                                            & Protocol                                                                & Attacks                                                                                                                                                                                                                                                                      & \begin{tabular}[c]{@{}l@{}}Attack \\ Detection\end{tabular} & \begin{tabular}[c]{@{}l@{}}Attack \\ Prevention\end{tabular} & Approach                                                                                                                                                                                                                                                                                                                          & \begin{tabular}[c]{@{}l@{}}Protocol \\ Inspection\end{tabular} & NIDS \\ \hline
1                                                 & \cite{yang2016multidimensional} & \begin{tabular}[c]{@{}l@{}}500kV smart\\ substation\\ \cite{7286357} with IEDs, \\ switches, \\ and SCADA\end{tabular}                                                                                                               & \begin{tabular}[c]{@{}l@{}}pcap:\\ Benign \& \\ Attack\end{tabular}            & \begin{tabular}[c]{@{}l@{}}IEC61850:\\ MMS,\\ GOOSE,\\ SMV\end{tabular} & \begin{tabular}[c]{@{}l@{}}318 types of\\ abnormal packets \\ generated by modifying \\ captured packets \\ and malicious \\ packets injection\end{tabular}                                                                                                                    & Y                                                           & Y                                                            & \begin{tabular}[c]{@{}l@{}}Access-Control,\\ Protocol Whitelisting,\\ Model-based detection,\\ Multiparameter based \\ detection\end{tabular}                                                                                                                                                                                     & N                                                                     & Y    \\ \hline
2                                                 & \cite{de2025rnn}                & \begin{tabular}[c]{@{}l@{}}RTDS + IED \\ HIL setup\end{tabular}                                                                                                                  & \begin{tabular}[c]{@{}l@{}}csv:\\ Benign \& \\ Attack\end{tabular} & GOOSE                                                                   & \begin{tabular}[c]{@{}l@{}}masquerade,\\ replay,\\ message injection,\\ poisoning attacks.\end{tabular}                                                                                                                                                                       & Y                                                           & N                                                            & \begin{tabular}[c]{@{}l@{}}RNN\end{tabular}                                                                                                                                                                                                                                                       & N                                                                     & N    \\ \hline
3                                                 & \cite{jay2025intelligent}       & \begin{tabular}[c]{@{}l@{}}Testbed for\\ PGCIL and\\ IEC61850Security\\Dataset \cite{8909783}\end{tabular}                                                                        & \begin{tabular}[c]{@{}l@{}}pcap:\\ Benign \&\\ Attack\end{tabular}            & GOOSE                                                                   & \begin{tabular}[c]{@{}l@{}}Data Manipulation,\\ MITM,\\ replay,\\ DoS\end{tabular}                                                                                                                              & Y                                                           & N                                                            & \begin{tabular}[c]{@{}l@{}}Data-driven:\\ (statistical + ensemble\\ unsupervised learning)\\ Model-based:\\ using SLD.\end{tabular}                                                                                                                   & N                                                                     & N    \\ \hline
4                                                 & \cite{baranwal2025domain}       & \begin{tabular}[c]{@{}l@{}}IEDs, IDS, HMI,\\ an attacker PC.\end{tabular}                                                                                                                                                               & \begin{tabular}[c]{@{}l@{}}pcap:\\ Benign \& \\ Attack\end{tabular}                                                                           & MMS                                                                     & \begin{tabular}[c]{@{}l@{}} Scanning IED\\ structure and IED,\\ invalid journal read\end{tabular}                                                                                                                                       & Y                                                           & Y                                                            & \begin{tabular}[c]{@{}l@{}}Rule based IDS\\ and ML-IDS\\anomaly detection\\ using domainId \&\\ itemID\end{tabular}                                                                                                                                       &  Y  & Y    \\ \hline
5                                                 & \cite{yadgar2025detection}     & EPIC (8 scenarios)                                                                                                                                                                                                                                          & \begin{tabular}[c]{@{}l@{}}pcap:\\ Benign\end{tabular}                     & MMS                                                          & \begin{tabular}[c]{@{}l@{}}DoS, FDI,\\ PLC code\\ manipulation\end{tabular}                                                                                                                                                                     & Y                                                           & N                                                            & \begin{tabular}[c]{@{}l@{}}sensor limits,\\ signal behavior,\\ physical power\\ system models, \\ and PCA\end{tabular}                                                                                                                                     & N                                                                     & N    \\ \hline
6                                                 & \cite{baranwal2025critical}     & \begin{tabular}[c]{@{}l@{}}Two IEDs,\\ Raspberry Pi,\\ Rule-based IDS,\\ Windows \\attacker PC\end{tabular} & \begin{tabular}[c]{@{}l@{}}pcap:\\ Benign \&\\ Attack\end{tabular}            & MMS                                                                     & \begin{tabular}[c]{@{}l@{}}IEDExplorer scanning,\\ spoofing protocol fields\end{tabular}                                                                                                                            & Y                                                           & N                                                            & \begin{tabular}[c]{@{}l@{}}OCC detection,\\ MMS Error-\\Aware IDS\end{tabular}                                                                                                                                                                             & Y                                                                     & Y    \\ \hline
7                                                 & \cite{albarakati2021security}   & \begin{tabular}[c]{@{}l@{}}Cosimulation \\testbed: OPAL-RT \\+ OpenStack\end{tabular}                                                                                                                                                 & \begin{tabular}[c]{@{}l@{}}pcap:\\ Benign\end{tabular}                                                                           & GOOSE                                                                   & \begin{tabular}[c]{@{}l@{}}Malicious Packets\\ Injection and\\ Delaying legitimate\\ packets.\end{tabular}                                                                                                                                                                 & Y                                                           & N                                                            & \begin{tabular}[c]{@{}l@{}}LSTM, GRU, RNN \end{tabular}                                                                                                                            & Y                                                                     & N    \\ \hline
8                                                 & \cite{yang2022new}              & \begin{tabular}[c]{@{}l@{}}IEC61850 smart\\ substation\\ (China Govt.)\end{tabular}                                                                                                                                                                   & \begin{tabular}[c]{@{}l@{}}pcap:\\ Benign \&\\ Attack\end{tabular}            & \begin{tabular}[c]{@{}l@{}}GOOSE,\\ MMS\end{tabular}                                                           & \begin{tabular}[c]{@{}l@{}}TCP-SYN Flood,\\ Land attack,\\ Smurf attack, \\ FTP attack, \\ GOOSE replay attack, \\ GOOSE attack with\\ high “stNum”, \\ MMS malformed packet, \\ MMS\_request and \\ MMS\_confirm\\ Flooding attack\end{tabular} & Y                                                           & N                                                            & LSTM-Autoencoder                                                                                                                                                                                                                                               & N                                                                     & N    \\ \hline
9                                                 & \cite{zhu2020intrusion}         & \begin{tabular}[c]{@{}l@{}}IEEE 39-bus\\ test system\end{tabular}                                                                                                                                                                                                                                     & \begin{tabular}[c]{@{}l@{}}pcap:\\ Benign \&\\ Attack\end{tabular}                                                                           & MMS                                                                     & \begin{tabular}[c]{@{}l@{}}MMS measurement\\ attacks (single/\\multiple substations)\end{tabular}                                                                                                                                                                             & Y                                                           & Y                                                            & \begin{tabular}[c]{@{}l@{}}Substation level\\ distributed IDS\end{tabular} & N                                                                     & N    \\ \hline

10
& \cite{wang2020detection}         & \begin{tabular}[c]{@{}l@{}}IEEE 8-bus\\ \& IEEE 118-bus\\ smart grid \\simulation\end{tabular}                                                                                                                                                                                                                                     & NA                                                                           & NA                                                                     & FDI                                                                                                                                                                             & Y                                                           & N                                                            & \begin{tabular}[c]{@{}l@{}}Interval residuals\\-based detection\end{tabular} & N                                                                     & N    \\ \hline

11
& \cite{yang2016intrusion}         & \begin{tabular}[c]{@{}l@{}}500kV smart\\
substation\end{tabular}                                                                                                                                                                                                                                     & \begin{tabular}[c]{@{}l@{}}pcap:\\ Benign \&\\ Attack\end{tabular}                                                                           & \begin{tabular}[c]{@{}l@{}}GOOSE\\MMS\\SMV \end{tabular}                                                                   & \begin{tabular}[c]{@{}l@{}}unauthorized connections,\\protocol misuse, etc.\end{tabular}                                                                                                                                                                             & Y                                                           & N                                                            & \begin{tabular}[c]{@{}l@{}}Access,protocol\\ semantics based\\ anomaly detection\end{tabular} & Y                                                                     & Y    \\ \hline

12
& \cite{hong2014integrated}         & \begin{tabular}[c]{@{}l@{}}WSU substation\\testbed\end{tabular}                                                                                                                                                                                                                                     & \begin{tabular}[c]{@{}l@{}}pcap:\\ Benign \&\\ Attack\end{tabular}                                                                           & \begin{tabular}[c]{@{}l@{}}GOOSE\\SMV \end{tabular}                                                                    & \begin{tabular}[c]{@{}l@{}}replay,\\ packet modification,\\ injection,\\ generation\\ and DoS\end{tabular}                                                                                                                                                                             & Y                                                           & Y                                                            & \begin{tabular}[c]{@{}l@{}}Host \& Network\\ based Anomaly\\ Detection\end{tabular} & Y                                                                     & Y    \\ \hline
\end{tabular}
}
\end{table}

\section{System model and Proposed Framework} 
\label{sec:sys-model}
\subsection{System Model for Defending Power Grid}

\begin{figure}[htbp]
{\includegraphics[width=\columnwidth]{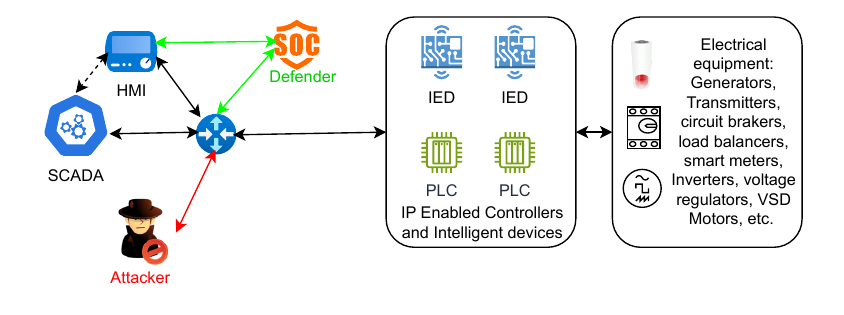}}
\caption{System Model}
\Description{System Model}
\label{fig:SystemModel}
\end{figure}

The smart grid represents a modernized power system that relies on layered monitoring, communication, and coordinated control across various components, including generators, transformers, circuit breakers, smart meters, inverters, variable-speed drives, and other field-level electrical equipment. Figure~\ref{fig:SystemModel} presents the system model of a layered smart-grid control system, illustrating how supervisory control systems, intelligent electronic devices, programmable controllers, communication networks, and field equipments interact to enable real-time monitoring, decision-making, and intelligent operation of the overall system.

The supervisory layer, includes the SCADA system and the Human–Machine Interface (HMI) responsible for system monitoring, visualization, alarm handling, and issuing control commands; the control and protection layer, consists of Programmable Logic Controllers (PLCs) and Intelligent Electronic Devices (IEDs), collectively referred to as control and protection units, that perform protection logic, equipment control, and real-time data acquisition from electrical assets such as circuit breakers, transformers, voltage regulators, and inverters; and the field layer, represents the actual electrical equipment executing physical operations which includes devices such as generators, transformers, circuit breakers, disconnect switches, capacitor banks, voltage regulators, smart meters, variable-speed drives, and power-electronic converters. These components are interconnected through an industrial Ethernet network that enables bidirectional data and command flow between SCADA, HMI, and control units. 

In operation, electrical field measurements are obtained at the source by instrument transformers, such as current transformers (CTs) and voltage transformers (VTs) convert high line current and voltage into proportionally reduced and safe secondary values that reflect the true electrical conditions on the primary system; CTs produce a scaled current output (typically 1 A or 5 A) proportional to the actual line current while isolating the high-voltage circuits, and VTs produce a proportionally reduced voltage output that reflects the system voltage for further processing by downstream devices, ensuring accurate and safe measurement without physical exposure to system voltages~\cite{Sheikh2025InstrumentTransformers}. 

These analog outputs (which can be current or voltage signals) are either directly hardwired as analog inputs to protection and control equipment or are digitized locally by processes such as merging units or analog-to-digital converters; in traditional substations CT/VT secondaries feed IEDs and PLCs with analog signals (e.g., proportional AC current/voltage levels), whereas in modern digital substations sampled values are digitized and communicated over digital process buses compliant with standards like IEC 61850 to protective and control devices~\cite{EEP2025VoltageCurrentMeasurement, EEP2024CentralizedProtectionControl}.

The real-time measurements of current and voltage originating from instrument transformers (CTs and VTs/PTs), together with frequency and power quantities (active kW, reactive kVAR, apparent kVA) computed from these signals by protection relays or power meters, and accumulated energy quantities obtained from dedicated metering devices, are then consumed by Intelligent Electronic Devices (IEDs), which are microprocessor-based controllers that analyze incoming measurement data and execute high-speed protection algorithms such as overcurrent, differential, distance, and voltage/frequency protection, and issue immediate trip commands to circuit breakers or other protective actuators to isolate faults and record event details.

In parallel, these measured values and equipment statuses are communicated to Programmable Logic Controllers (PLCs), which serve as automation and control processors that interpret sensor and IED data (typically via digital communication or I/O modules), execute logic-based sequences, regulate setpoints, control switching of motors, capacitor banks, or voltage regulators, and coordinate larger control actions such as load shedding or sequence control. The analog signals from CTs and VTs are interfaced to PLC and IED input modules where they are conditioned and, if necessary, converted to digital values for processing, while outputs from PLCs and IEDs to actuators such as circuit breakers, contactors, motor starters, tap changers, and power electronic converters may be analog or digital depending on the device and network architecture; PLCs typically issue logic level or fieldbus commands to actuators to execute control actions, whereas IEDs issue trip or block signals to breaker trip/close circuits or other protection equipment~\cite{EPRIIntelliGrid_DACUseCase}. 

The cooperation between IEDs and PLCs occurs through exchange of measurement, status, alarm, and command information: IEDs inform PLCs of fault conditions and protective states, enabling coordinated control responses such as feeder reconfiguration or system restoration, and PLCs provide permissive, blocking, or coordination signals to IEDs during planned switching or automation sequences to optimize overall system performance and maintain stability, forming a hierarchical flow from sensors → measurement conversion → protection decisions → control logic → actuators that ensures both fast fault isolation and stable system operation~\cite{Midence2022DigitalizationDisturbance}.

The SCADA system collects processed data from both PLCs and IEDs, updates visualizations on the HMI, and generates alarms when abnormal conditions occur. The human operator interacts with the SCADA/HMI interface to supervise system status, acknowledge alarms, review trends, and issue high-level control commands when necessary. While most routine actions are automated by the control and protection layer, the operator plays a critical role in decision-making during abnormal events, overseeing system performance, and initiating manual control actions during emergencies or maintenance activities.

Building on this layered architecture, our Defender can be seamlessly integrated at the network interface between the supervisory layer and the control/protection units. 

This placement aligns directly with the OT-specific guidance in NIST SP 800-82r3~\cite{stouffer2023nist} (document by NIST that provides guidance for establishing secure operational technology), which emphasizes that OT network traffic is typically deterministic, repeatable, and predictable. The standard notes that organizations should first develop an understanding of the normal state of OT network communication and that passive, listen-mode monitoring is often a necessary initial step for differentiating legitimate operational behavior from abnormal or malicious activity. 
It further highlights that encrypted communication may limit the visibility of IDS/BAD systems, and recommends adjusting the monitoring point, such as capturing traffic before or after encryption, to improve detection accuracy. Additionally, the standard states that IDS and IPS technologies are effective for detecting known internet attacks, and that many IDS and IPS vendors now provide OT-protocol-aware signatures for Modbus, DNP3, ICCP, and similar industrial protocols. It also recommends combining host-based and network-based monitoring for an effective deployment. 

In line with this guidance, our Defender module leverages the deterministic communication patterns of SCADA with PLCs and IEDs to identify deviations from authorized behavior, perform protocol-specific inspection, and support real-time threat mitigation within the substation automation network. 
Hence, by positioning our Defender module between supervisory and protection and control unit as shown in Figure~\ref{fig:SystemModel}, we provide real-time monitoring, anomaly detection, and rapid mitigation of malicious traffic targeting IP-addressable devices i.e PLCs and IEDs. This integration helps multiple stakeholders: operators receive immediate alerts on the HMI, enabling informed operational decisions; SOC analysts gain detailed traffic insights and adaptive rules for threat containment; and decision-makers benefit from enhanced situational awareness and system resilience. In essence, our Defender strengthens the smart-grid control system by ensuring both cybersecurity and operational reliability, complementing the automated control and protection functions already managed by PLCs and IEDs.

\subsection{Threat Model}
\subsubsection{Attacker Objective}
The attacker aims to conduct precise, protocol-aware manipulation within a industrial control environment. After gaining a foothold in the SCADA network, the adversary seeks to observe system state, understand operational behavior, and ultimately issue targeted control-oriented write operations that could alter process values or trigger switching actions. The objective is not blind disruption but informed, high-fidelity manipulation of application-layer semantics that convincingly mimics legitimate supervisory activity.  For instance, an attacker using an open-source industrial-protocol toolkit like iec61850bean tool, could identify which logical node and data object and attributes correspond to a circuit breaker’s operational state and send a packet that imitates a legitimate supervisory command aimed at changing it's state. 

\subsubsection{Attacker Knowledge}
The attacker is assumed to possess detailed familiarity with industrial communication protocols used in smart grid and IEC61850 data models used by field devices (IEDs and PLCs) and supervisory systems. Attacker understands how device parameters, operational attributes, and control interactions are structured and executed in practice. They are also aware of how commonly available protocol libraries and tooling implement these communication stacks and how such stacks typically operate within industrial control environments. This knowledge allows the attacker to interpret observed process variables and behavioral patterns, relate them to meaningful operational functions, and identify opportunities where protocol semantics could be misused in ways that are not apparent from basic network-layer monitoring, where inspecting headers alone is not sufficient to reveal deeper protocol misuse. For example, an attacker familiar with tools like iec61850bean might load a device’s data model to understand how its logical components relate to real-world equipment, enabling them to interpret which variables represent status indicators, measurements, or control targets. 

\subsubsection{Attacker Capability}
The attacker is a well-resourced insider or a lateral-movement adversary who has already obtained access to the industrial control network through some compromised host (for example, wireless AP). Using publicly available protocol toolkits, the adversary can rapidly enumerate endpoints, inspect protocol-level data models, observe live values, and generate protocol-aware operations at scale. These capabilities allow them to craft traffic that resembles legitimate supervisory interactions rather than relying on indiscriminate or blind techniques. This sophistication highlights the need for defensive mechanisms that detect misuse through protocol-level behavior and semantic anomalies. 
For example, by leveraging a protocol library like \emph{libiec61850} or tools such as \emph{iec61850bean}  that automatically parse logical nodes and attributes in a smart substation, an attacker could generate traffic that mirrors the structure of routine supervisory messages while gathering crucial information for more targeted cyber-physical disruptions. In such a scenario, an attacker’s capabilities are as follows, but not limited to:
\begin{itemize}
    \item Read Device Information (LPHD) to assess health, operational mode, and condition of the physical device.
    \item Read Logical Node Zero (LLN0) values to understand the overarching data model, including datasets, reporting structures, and enabled functions.
    \item Read Protection (PTOC, PDIS, PDIR, PTOV, PTUV) values to observe protection element states and thresholds relevant to system behavior.
    \item Read Circuit Breaker (XCBR) values to determine present breaker status and operational counters.
    \item Change Interlocking (CILO) state to affect permissive or blocking conditions for switching operations.
    \item Change Switch Controller (CSWI) state to issue switch open or close actions.
    \item Change Circuit Breaker (XCBR) state to drive the breaker into an open or closed condition.
    \item Change Measurement (MMXU) state to alter power, voltage, current, or frequency inputs used by control logic.
    \item Change Sequence and Imbalance (MSQI) state to modify phase-sequence or imbalance conditions.
    \item Change Metering (MMTR) state to impact accumulated energy or demand values.
    \item Change Overvoltage (PTOV) state to modify how the system responds to high-voltage conditions.
    \item Change Directional Overpower (PDOP) state to affect decisions based on overpower flow direction.
\end{itemize}

\subsection{Defender Model}

The Defender module (SOC/IDS node), as illustrated in Figure~\ref{fig:SystemModel}, is strategically positioned at the network interface connecting the supervisory layer to the control and protection units. It operates as a intrusion detection and prevention system, continuously monitoring traffic to and from these devices. A Network Intrusion Detection System (NIDS), equipped with rule sets derived from known protocol-specific attack signatures, filters and drops malicious packets in real time, while simultaneously capturing all communication traffic for in-depth protocol analysis. The Defender inspects protocol-specific packets, notably MMS read/write operations, to detect deviations from authorized behaviors. It verifies whether the field–value pairs that express the intent and operation of incoming packets, align with whitelisted references originating from approved devices such as the SCADA. Any packet whose operational semantics fall outside these approved field-value pairs is classified as a potential attack signature. In such cases, first the Defender traces the attack path i.e source IP (originator), destination IP (targetted PLC/IED), operation (mms service) and attacked process (though domainID and itemID), second raises an alert on the HMI, and automatically generates a new NIDS rule to block similar packets in the future. Leveraging the fact that PLCs and IEDs are IP-addressable endpoints with fixed roles and deterministic communication patterns, the Defender can precisely attribute traffic to specific devices, extract repeatable operation signatures, identify malicious control attempts, and enforce rapid containment. Our Defender Model enables proactive detection, adaptive rule generation, and real-time mitigation of cyber intrusions targeting substation automation networks.

\section{Proposed Framework}

\begin{figure*}[htbp]
{\includegraphics[width=\columnwidth]{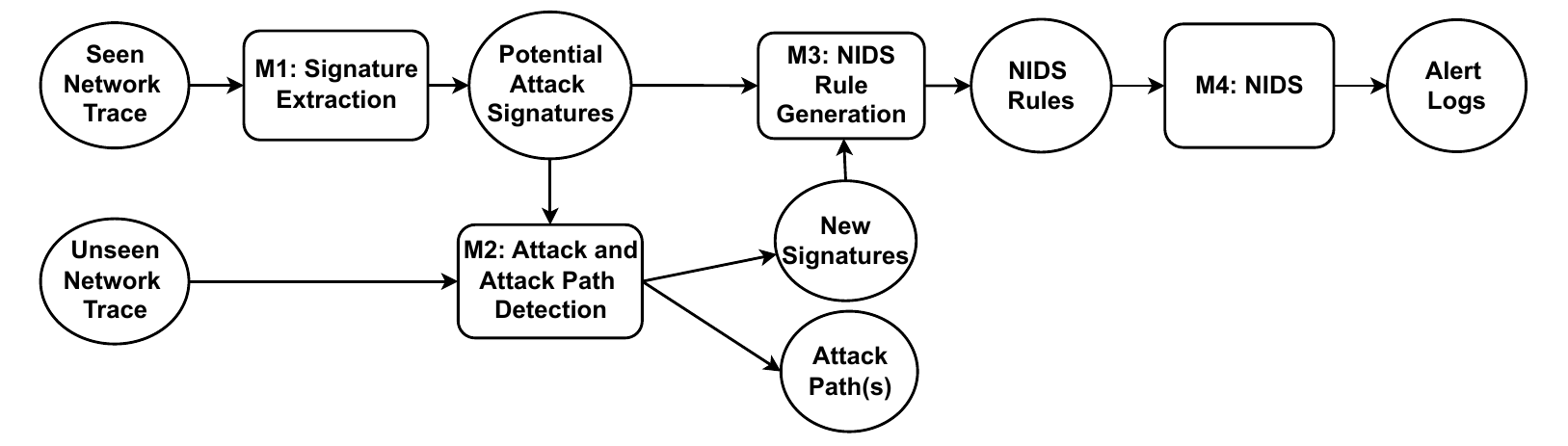}}
\caption{Overview of modules in proposed framework. M1: Signature Extraction, M2: Attack and Attack Path Detection, M3: NIDS Rule Generation and M4: NIDS}
\label{fig:framework}
\end{figure*}
Existing research has analyzed various MMS-PDU header fields, such as mms.confirmedServiceRequest, mms.listOfIdentifier, and mms.getVariableAccessAttributes, to detect reconnaissance activities, including IED scanning and invalid journal read attempts. These approaches typically rely on identifying abnormal packet volumes or inconsistencies in fields like domainId and itemId. 
Also SCD files and IEC 61850 traffic profiles are used to model normal behavior, the SCD file, as per IEC 61850-6, captures only design-time IED configuration and IED-to-IED communications, not SCADA-to-IED MMS interactions. Because these configuration views cannot represent runtime SCADA behavior, they are insufficient for detecting attacks that mimic legitimate MMS operations.

Also, history has shown through incidents such as Industroyer that attacks on power systems can be far more subtle and unpredictable. Malicious actions may take the form of legitimate looking read or write operations issued through authorized entities, such as SCADA or HMIs, and cause disruptive effects on the underlying control infrastructure.

In our proposed framework, we move beyond traditional IP-based identification and instead focus on MMS-specific header and payload features to distinguish between normal and malicious activity. Our approach relies solely on inspecting network traffic to extract attack signatures and detect them in real time. We believe, whitelisting derived from real MMS network traffic, captures the true operational behavior of SCADA–IED/PLC communication and enables the detection of unauthorized read/write actions.

Overall, we present a framework capable of identifying unauthorized read and write operations within MMS communication, which is commonly used between SCADA systems and IEDs/PLCs. This enables the detection of both overt and stealthy attacks that exploit legitimate communication channels.
 
Our proposed framework as shown in Figure~\ref{fig:framework} has four modules: M1: Signature Extraction, M2: Attack and Attack Path Detection, M3: NIDS Rule Generation and M4: NIDS. 
Module M1 inspects both MMS benign and attack traffic. It extracts attack signatures for both MMS read and write operations using a whitelisting-based method. Specifically for MMS write operations it also detects abnormal values in specific fields (identified and discussed in~\ref{M1}) of the MMS-PDU. 
Module M2 applies detected attack signatures from M1 on real time MMS traffic and detects malicious packets. Also in parallel detects fields-value pairs in MMS-PDU that are neither part of whitelist nor listed in identified attack signatures and identifies them as potential new attack signatures.
M1 and M2 together, can provide both precise detection of known malicious behaviors and robust identification of deviations that may indicate previously unseen threats.
All the extracted and newly detected attack signatures from M1 and M2 modules respectively, are fed to M3 module to automatically generate rules for NIDS. The M4 module comprises the deployed NIDS which takes a rule file and applies to the incoming MMS packets. 

\subsection{M1: Signature Extraction}
\label{M1}
We assume that a SCADA read or write operation is normal and genuine only when it results from valid operator intervention, namely, when read operations occur for manual verification, troubleshooting, or process assessment, and when write operations occur for intentional control actions, setpoint changes, procedural steps, or authorized overrides of automatic logic. Conversely, we assume an attack scenario when read or write operations originate from external tools or scripts, such as those built using IEC61850Bean, libiec61850, or similar libraries, rather than from authenticated SCADA interface.

To extract attack signatures for both MMS read and write operations, we follow a structured three-step process. 

First, from benign MMS traffic, we extract MMS fields ip.src, ip.dst, mms.confirmedServiceRequest, mms.domainId, mms.itemId, mms.iec61850.timeAccuracy, and mms.data.octetString and then separate packets into read and write operations based on request type identification using \emph{mms.confirmedRequestService} field value. We build two whitelists: a read whitelist containing legitimate (mms.domainId, mms.itemId) pairs, and a write whitelist containing permissible (mms.domainId, mms.itemId, mms.iec61850.timeAccuracy, mms.data.octetString) tuples.

Next, from the attack traffic, we again extract MMS fields and categorize packets using request type identification and form potential signature lists of read pairs and write tuples. By comparing these with their respective whitelists, we isolate only those elements that appear exclusively in the attack traffic, these constitute the potential attack signature list.

Finally, each signature in the potential attack signature list is examined to determine whether it corresponds to an operation targeting a substation component. If so, it is classified as a valid attack signature; otherwise, it is discarded as non-malicious. 


\begin{figure*}[htbp]
{\includegraphics[width=\columnwidth]{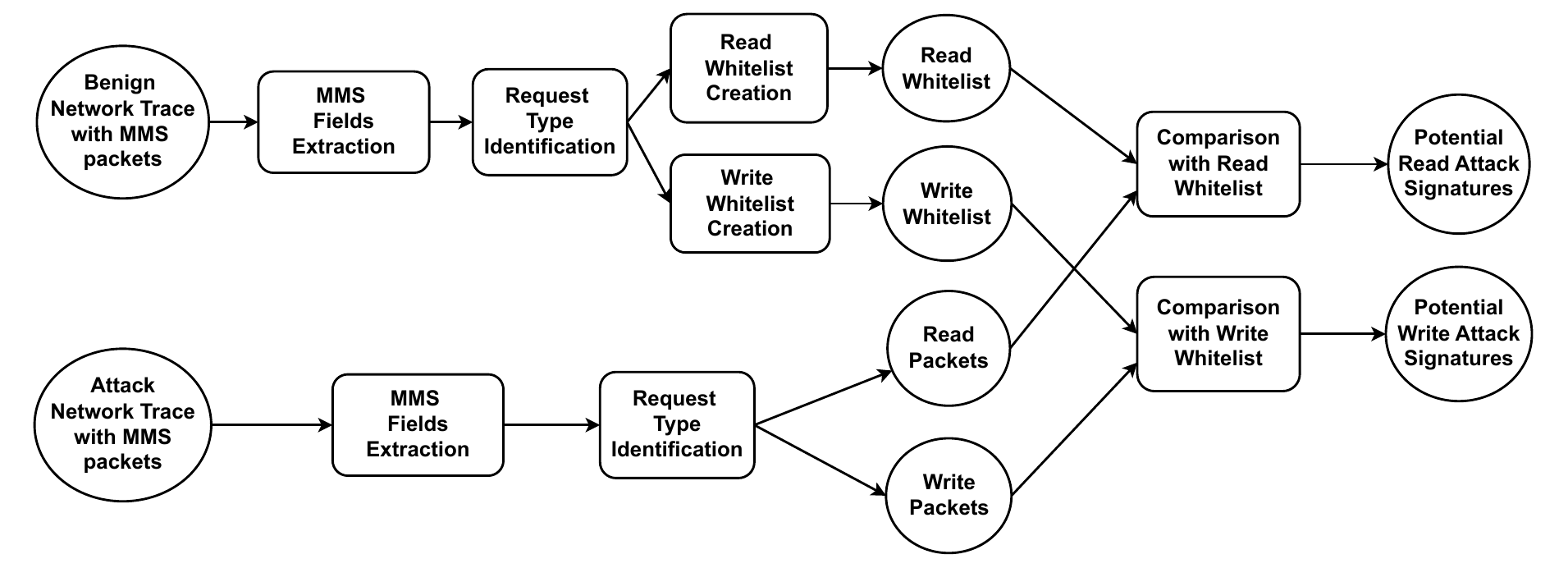}}
\caption{Steps in M1: Signature Extraction.}
\label{fig:M1framework}
\end{figure*}

\subsection{M2: Attack Path Detection}
Attack path detection module takes previously unseen network traffic containing MMS packets as input and extracts the mentioned MMS protocol fields. It then applies the whitelists and attack signatures generated by the M1 module to the incoming data. When a match with an existing attack signature is identified, the module reconstructs and outputs the corresponding attack path. Conversely, if the extracted fields do not align with any known attack signatures and are absent from the established whitelist, they are identified as potential new attack signatures. 

\begin{figure*}[htbp]
{\includegraphics[width=\columnwidth]{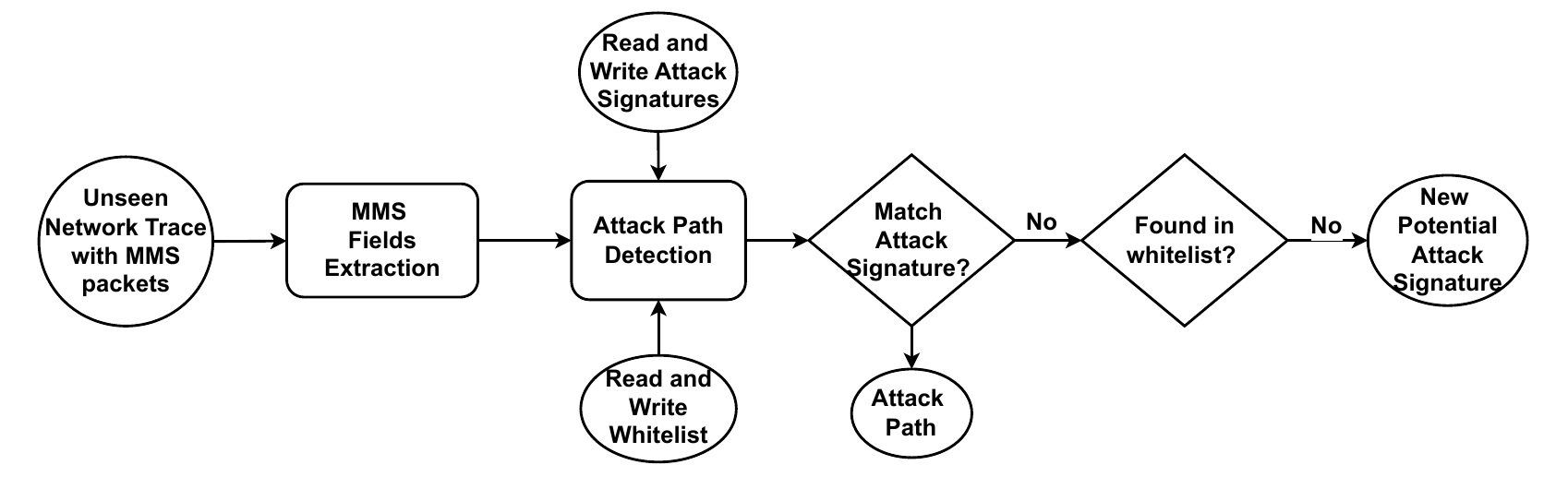}}
\caption{Steps in M2: Attack path Detection.}
\label{fig:M2framework}
\end{figure*}

\subsection{MMS (Manufacturing Message Specification)}



\begin{figure}[ht]
    \centering
    \begin{minipage}[t]{0.47\textwidth}
        \centering
        \includegraphics[width=\textwidth]{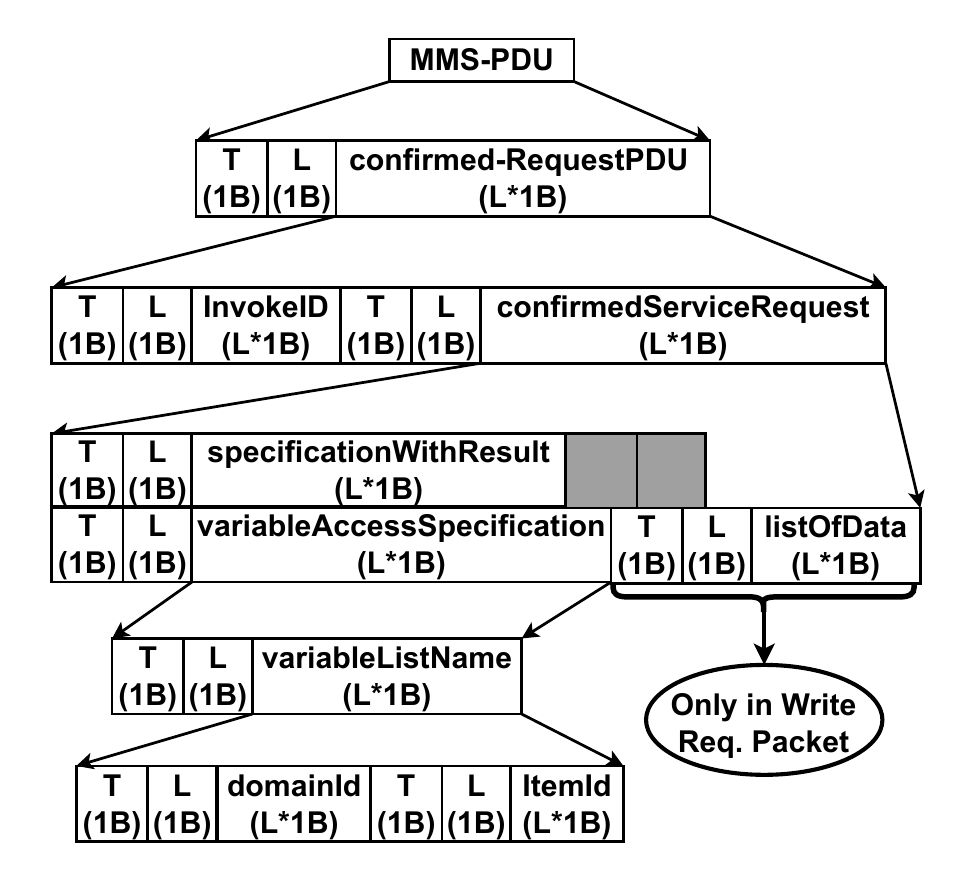}
        \caption{MMS Read/Write Request Packet Structure. (T: Tag (type), L: Length as per the BER-TLV structure) Optionally the L fields can precede 1B field with value 0x81 which indicates that content length in the value part is greater than 127B.}
        \Description{MMS Read/Write Request Packet Structure}
        \label{fig:MMS_readreq}
    \end{minipage}\hfill
    \begin{minipage}[t]{0.47\textwidth}
        \centering
        \includegraphics[width=\textwidth]{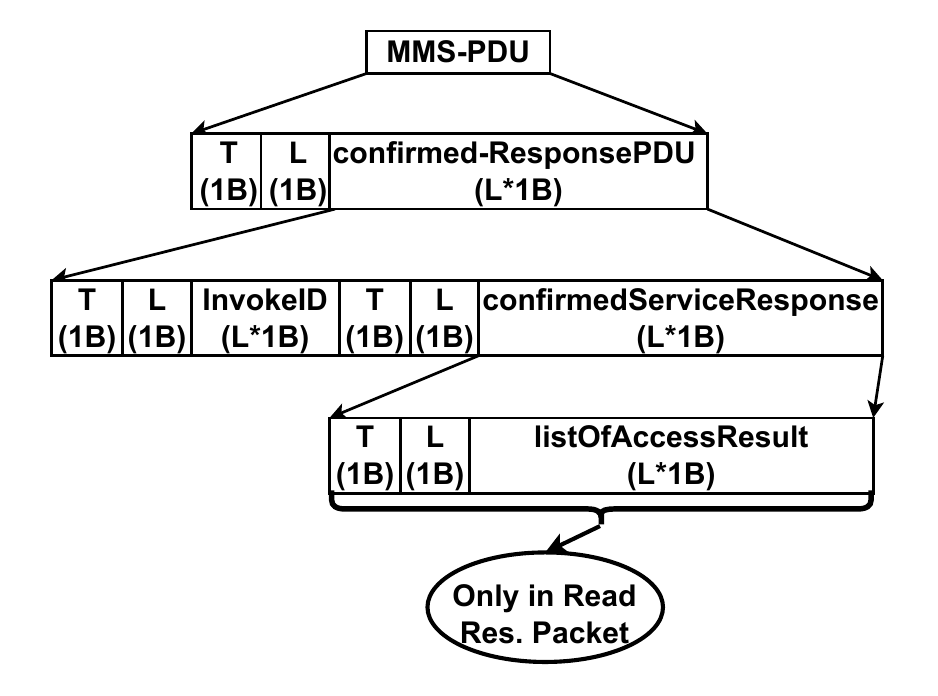}
        \caption{MMS Read/Write Response Packet Structure.}
        \Description{MMS Read/Write Response Packet Structure.}
        \label{fig:MMS_readres}
    \end{minipage}
\end{figure}

The IEC 61850 standard, defines the MMS (Manufacturing Message Specification) protocol as a client/server communication model used to exchange non-critical information over wide geographical areas, between Intelligent Electronic Devices (IEDs) and higher-level systems such as SCADA over Ethernet. MMS operates on top of the TCP/IP stack, allowing clients (SCADA) to connect to servers (PLC/IEDs) via IP address to read and write data, retrieve configurations, and exchange files. An IED acts as the MMS server, while the SCADA system functions as the client. The IED stores information in its data sets (such as actions, triggers, or position feedback) and automatically reports these values to the MMS client whenever a pre-defined trigger condition is met.

IEC 61850 needs a way to carry its abstract data models and services across a network, hence IEC 61850 maps its client-server services onto MMS, as MMS can easily support the complex naming and service models of IEC 61850~\cite{mackiewicz2006technical}. IEC 61850 defines what data looks like and what you can do with it, whereas MMS defines how those requests and responses are actually sent over the wire.

In the IEC 61850 standard, MMS protocol provides access to the internal data structure of Intelligent Electronic Devices (IEDs) through a well-defined hierarchy of named variables. These variables are derived from the object-oriented model of IEC 61850, in which each physical device comprises one or more logical devices (LDs), each containing logical nodes (LNs) that represent specific functions of the IED. These logical nodes consist of standardized data objects (DOs), which in turn contain data attributes (DAs), forming a hierarchical tree structure as shown in Figure~\ref{fig:data_model}. The naming and access methods for each object are standardized. 


\begin{minipage}[c]{\textwidth}
  \begin{minipage}[c]{0.495\textwidth}
    \centering
\includegraphics[width=0.9\textwidth]{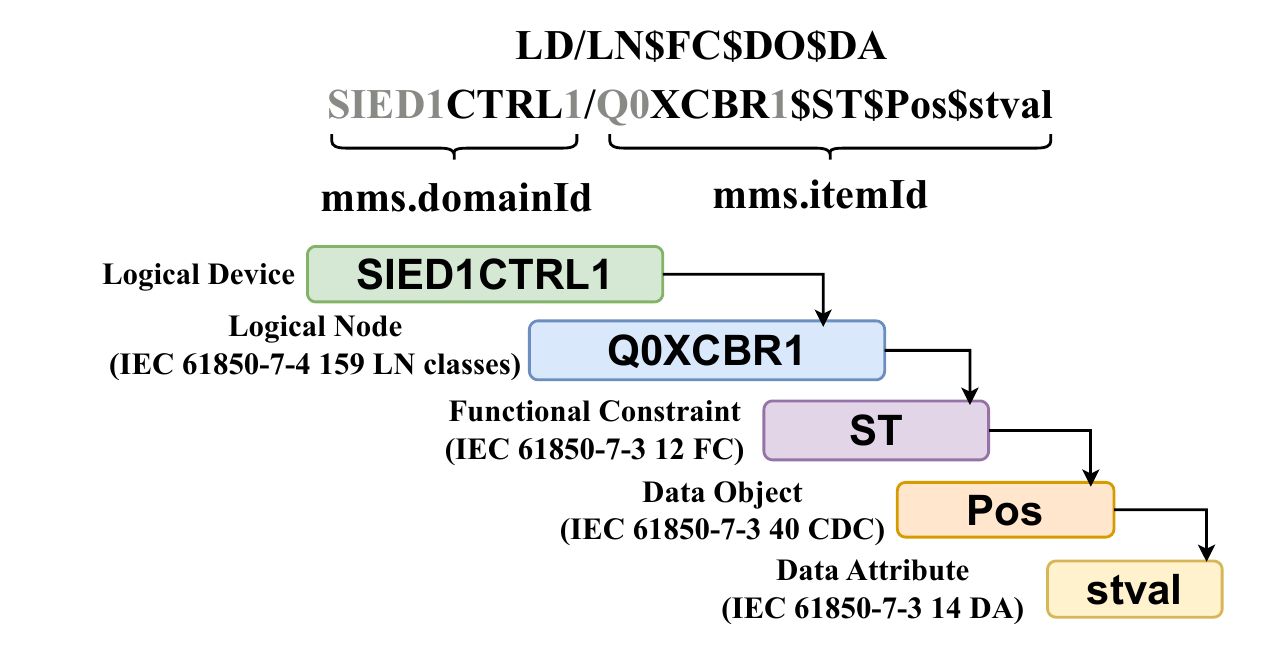}
\captionof{figure}{IEC61850 Data Model.}
  \Description{IEC61850 Data Model.}
\label{fig:data_model}
  \end{minipage}\hspace{0.0in}
  \begin{minipage}[c]{0.45\textwidth}
  \footnotesize
    \centering
    \captionof{table}{MMS Confirmed Services~\cite{falk1996mms, matouvsek2018description}}
\label{tab:MMS_services}
\scalebox{0.9}{
\begin{tabular}{|l|l|l|}
\hline
\textbf{\begin{tabular}[c]{@{}l@{}}MMS\\ Service\end{tabular}}            & \textbf{\begin{tabular}[c]{@{}l@{}}Tag \\ (dec)\end{tabular}} & \textbf{Description}                                                                                                                               \\ \hline
getNameList                                                               & 1                                                             & \begin{tabular}[c]{@{}l@{}}Requests logical node\\ name of the physical device\end{tabular}                                                        \\ \hline
read                                                                      & 4                                                             & \begin{tabular}[c]{@{}l@{}}Reads value of a given\\ list of variables\end{tabular}                                                                 \\ \hline
write                                                                     & 5                                                             & \begin{tabular}[c]{@{}l@{}}The service sets values\\ of a given list of variables\end{tabular}                                                     \\ \hline
\begin{tabular}[c]{@{}l@{}}getVariable\\ AccessAttributes\end{tabular}    & 6                                                             & \begin{tabular}[c]{@{}l@{}}The request retrieves MMS type\\ specification for given domain and\\ object. Ex: name=stVal, type=boolean\end{tabular} \\ \hline
\begin{tabular}[c]{@{}l@{}}getNamedVariable\\ ListAttributes\end{tabular} & 12                                                            & \begin{tabular}[c]{@{}l@{}}Requests available attributes\\ (variables) in the given dataset\end{tabular}                                           \\ \hline
\end{tabular}
}
    \end{minipage}
  \end{minipage}

For example, an IED might have a logical device named LD1 (Ex: SIED1CTRL1, TIED2MEAS1, etc.), containing several logical nodes such as XCBR (circuit breaker), CSWI (switch controller), XSWI (circuit switch), and MMXU (measurements). The XCBR logical node includes a data object called Pos, which provides information about the breaker’s position, including attributes like stVal (state value), q (quality), and t (timestamp). Accessing specific data within this structure follows a defined path as shown in Figure~\ref{fig:data_model}; for instance, the state value of the breaker Q0XCBR1 would be accessed using the path SIED1CTRL1/Q0XCBR1\$ST\$Pos\$stVal, while accessing all attributes of the Pos object would use the path SIED1CTRL1/Q0XCBR1\$ST\$Pos.

 For example, a logical node representing a circuit breaker (XCBR) may contain a control-related data object (CO), which includes sub-objects such as Pos (position) and BlkOpn (block open). Each of these sub-objects contains multiple data attributes like ctlVal (control value), operTm (operation time), origin, and ctlNum (control number). Access to these elements is standardized through MMS named variable paths, which follow a hierarchical naming convention. For instance, a client can access the entire CO object, a specific data object such as Pos, or individual attributes like ctlVal or operTim using paths such as XCBR1\$CO, XCBR1\$CO\$Pos, or XCBR1\$CO\$Pos\$ctlVal, respectively. 
 
 These named variables exist within the context of a Virtual Manufacturing Device (VMD), a conceptual abstraction defined by the MMS protocol that represents the virtual interface of a physical device. The VMD encapsulates all accessible data and services of the device, enabling standardized read, write, and control operations by MMS clients. This structured approach ensures interoperability and efficient communication between IEDs and supervisory systems such as SCADA.

There are 79 MMS Confirmed services, few are as follows along with their TAGs (an ASN.1 element identifier that marks what kind of object, service, or data type is being carried in the MMS message) and encoded TAGs in Table~\ref{tab:MMS_services}.

\subsubsection{MMS Packet Structure}

Figure~\ref{fig:MMS_readreq} illustrates the Manufacturing Message Specification (MMS) Read/Write Request Packet Structure, which defines the formalized data encapsulation used for service requests in MMS-based communication systems. The packet structure adheres to the Basic Encoding Rules (BER) of the Abstract Syntax Notation One (ASN.1) format, which employs the Tag-Length-Value (TLV) convention for representing data elements in a hierarchically encoded form~\cite{sorensen2008analysis}.

Each field within the MMS request packet is represented in a Tag (T) – Length (L) – Value (V) format, with the Tag and Length indicating the type and size of the following Value. The tag field (1 byte) specifies the semantic meaning or data type, while the length field (1 byte) denotes the size of the value field. When the value exceeds 127 bytes, the encoding introduces an extension byte with a value of 0x81, indicating that the subsequent byte contains the actual content length. This ensures support for large data payloads while maintaining compatibility with compact message structures.

The MMS Protocol Data Unit (PDU) begins with the confirmed-RequestPDU, encapsulating the invokeID and confirmedServiceRequest elements. The invokeID uniquely identifies each service transaction, enabling consistent request–response matching. The confirmedServiceRequest element defines the specific service invocation (e.g., read or write operation), further includes the specificationWithResult and variableAccessSpecification fields. These define the access context of the operation, identifying target variables or data objects, through subfields such as variableListName, domainId, and itemId.

The terminal section of the structure, listOfData, carries the actual data values in write requests or serves as a placeholder for retrieved values in read operations.


\subsubsection{MMS DomainId and ItemId}
In the Manufacturing Message Specification (MMS) model, the Domain Id (mms.domainId) and Item Id (mms.itemId) together form the complete MMS object reference, which is used to uniquely identify data objects, attributes, or control points within an MMS server (IED/PLC). 
The Domain Id designates the logical device or domain, acting as a container for related data, while the Item Id specifies the exact data object and corresponding attributes inside that domain along with functional constraints. 
For example, 
domainId: SIED1CTRL itemId: BI6GGIO1\$ST\$Health, points to the health status of a logical node in the SIED1CTRL domain,
domainId: GIED2MEAS with itemId: MMXU1\$MX\$TotPF\$instMag\$f, refers to the instantaneous magnitude of the total power factor measurement in the GIED2MEAS domain. 
Together, they provide a globally unique reference to any accessible element in the MMS server’s information model.
This pair is very important to understand what component is being tagetted by a mms packet.

\section{Experimental Setup}
\subsection{EPIC Testbed}
\begin{figure*}[htbp]
\centering
{\includegraphics[width=1\columnwidth]{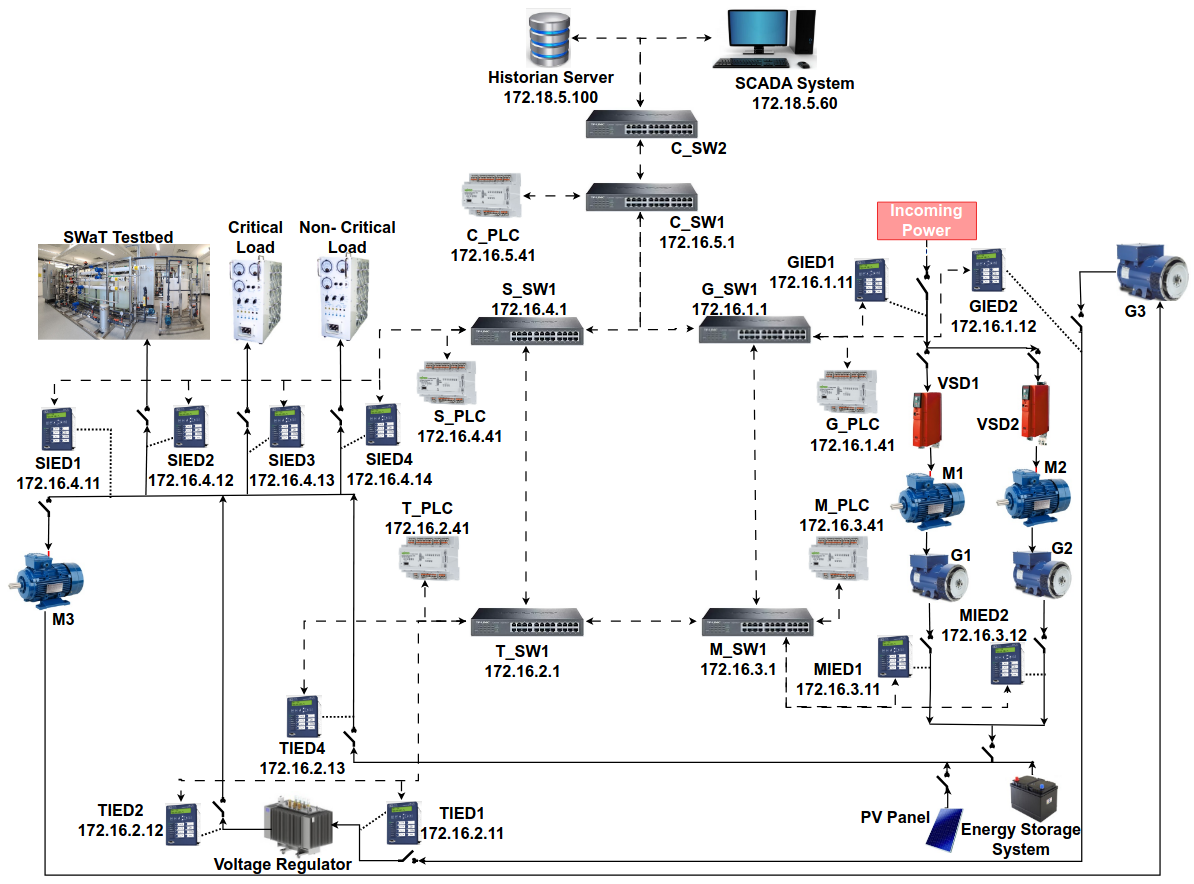}}
\caption{iTurst's EPIC Network Architecture with Physical Processes.}
\label{fig:EPIC_diagram}
\end{figure*}

The EPIC (Electric Power and Intelligent Control) testbed, developed by iTrust, is a scaled-down and fully operational replica of a modern smart power grid. EPIC is a unique testbed that facilitates realistic power grid operations, thereby enabling researchers to systematically design, evaluate, and validate defense mechanisms against both cyber and physical threats.
The EPIC testbed is divided into four stages: Generation, Transmission, Micro-grid, and Smart Home. The Generation stage features a power source from SUTD’s grid and two generators G1 and G2. The Micro-grid stage connects the generators, 110 photovoltaic cells (PV Panel), and batteries (Energy Storage Systems). In the Transmission stage, an autotransformer steps up the voltage for distribution. The Smart Home stage includes multiple loads such as critical and non-critical loads, iTrust’s SWaT testbed and a motor (M3).  
Figure~\ref{fig:EPIC_diagram} illustrates the physical processes across the four stages of the system.

\subsubsection{Communication Architecture}
WAGO PLCs control the opening and closing of breakers and loading within the EPIC testbed. Communication is organized into five subnets: Generation, Transmission, Micro-Grid, Smart Home, and SCADA. Each stage has its own switches, PLCs, power supply units, and protection systems, connected via a fiber-optic ring network to improve communication speeds. Redundancy is ensured through High-availability Seamless Redundancy (HSR) and Media Redundancy Protocol (MRP) switches. EPIC follows the IEC 61850 standard for communication in electrical substations and uses Generic Object Oriented Substation Event (GOOSE) and Manufacturing Message Specification (MMS) protocols for data exchange between relays and SCADA.

\subsection{EPIC Dataset collected on EPIC Testbed}

\begin{itemize}
    \item \textbf{[D1] \emph{Scenario (1...8)} Dataset}: This dataset can be accessed upon request from iTrust, Singapore. It has 8 folders, each having a pcap file \emph{(Scenario\_(1...8).pcapng)} and a csv file, collected over 30-minute intervals for eight different scenarios on the EPIC testbed. Scenario 1 involved synchronizing two generators (G1 and G2) without load, with the angle difference between them varying from -180 to 180 degrees. Scenario 2 included synchronization with a 10kW resistive load, again with the angle difference between G1 and G2 from -180 to 0 degrees. Scenario 3 involved two generators running with a 10kW resistive load. Scenario 4 and Scenario 5 tested two generators with a 10kW and 7kW resistive load, respectively, while also running a PV system. Scenario 6 involved three generators running with a 14kW resistive load. Scenario 7 and Scenario 8 focused on two generators supplying power to iTrust’s Secure Water Treatment (SWaT) testbed and both SWaT and Water Distribution (WADI) testbeds, respectively. The dataset includes MMS traffic between SCADA and PLC, as well as between SCADA and IED.

\begin{table}[]
\caption{Details of Scenario dataset under normal operations}
\label{tab:scenario_dataset}
\scalebox{0.9}{
\begin{tabular}{|l|l|l|l|l|}
\hline
\textbf{Filename}  & \textbf{\#packets} & \textbf{\begin{tabular}[c]{@{}l@{}}\#MMS \\ packets\end{tabular}} & \textbf{\begin{tabular}[c]{@{}l@{}}MMS \\ Service\\ Req.\end{tabular}}   & \textbf{\begin{tabular}[c]{@{}l@{}}MMS\\ Asso.\\ Req\end{tabular}} \\ \hline
Scenario\_1.pcapng & 449177             & 378265                                                            & \begin{tabular}[c]{@{}l@{}}4: 189070\\ 1: 2\\ 12: 8\\ 5: 49\end{tabular} & 1                                                                  \\ \hline
Scenario\_2.pcapng & 144291             & 121223                                                            & \begin{tabular}[c]{@{}l@{}}4: 60593\\ 5: 18\end{tabular}                 & 0                                                                  \\ \hline
scenario\_3.pcapng & 578150             & 486107                                                            & 4: 243053                                                                & 0                                                                  \\ \hline
scenario\_4.pcapng & 914714             & 769956                                                            & \begin{tabular}[c]{@{}l@{}}4: 384971\\ 5: 7\end{tabular}                 & 0                                                                  \\ \hline
scenario\_5.pcapng & 498615             & 418676                                                            & \begin{tabular}[c]{@{}l@{}}4: 209336\\ 5: 2\end{tabular}                 & 0                                                                  \\ \hline
scenario\_6.pcapng & 560312             & 472794                                                            & \begin{tabular}[c]{@{}l@{}}4: 236384\\ 5: 13\end{tabular}                & 0                                                                  \\ \hline
scenario\_7.pcapng & 652877             & 585898                                                            & 4: 292947                                                                & 0                                                                  \\ \hline
scenario\_8.pcapng & 1082881            & 970155                                                            & 4: 485078                                                                & 0                                                                  \\ \hline
\end{tabular}
}
\end{table}

    \item \textbf{[D2] \emph{Oct 2021} Dataset}: This dataset can be accessed upon request from iTrust, Singapore. It has 5 pcap files. This dataset represents power consumption measurements from a testbed consisting of three resistive load banks: an electronic load bank, a non-critical load bank, and a critical load bank. Each data point mentioned in the  \emph{Readme.txt}, has Time index, electronic bank (kilowatt), critical load bank (kilowatt) values (Eg. 20, 2.3*3, 0.5, 1;
21, 2,3*2, 0.5, 0.5;
22, 2.3*3, 0.5, 1;
23, 2.3*3, 0.5, 0;
24, 2.3*3, 1, 1;
25, 2.3*3, 5.5, 0.5;
26, 2.3*3, 5.5, 1;
27, 2.6*3, 5.5, 0.5) corresponds to a 15-minute interval over a 2-hour experimental window. The loads are recorded in kilowatts, with the electronic load operated in parallel channels (multiples of a base value). From time index 52 onward, the non-critical load bank was removed due to a fault, and only the electronic and critical load banks remained in operation. The data provides insight into dynamic load behavior under varied operational scenarios.
\begin{table}[]
\caption{Details of Oct 2021 dataset under normal operations}
\label{tab:Oct2021_dataset}
\scalebox{0.9}{
\begin{tabular}{|l|l|l|l|l|}
\hline
\textbf{Filename}                                                         & \textbf{\#packets} & \textbf{\begin{tabular}[c]{@{}l@{}}\#MMS \\ packets\end{tabular}} & \textbf{\begin{tabular}[c]{@{}l@{}}MMS \\ Service\\ Req.\end{tabular}} & \textbf{\begin{tabular}[c]{@{}l@{}}MMS\\ Asso.\\ Req\end{tabular}} \\ \hline
\begin{tabular}[c]{@{}l@{}}index20\_27\_1600\_\\ 1800.pcapng\end{tabular} & 8418641            & 5087164                                                           & \begin{tabular}[c]{@{}l@{}}4: 2543559\\ 5: 22\end{tabular}             & 0                                                                  \\ \hline
\begin{tabular}[c]{@{}l@{}}index28\_35\_1245\_\\ 1445.pcapng\end{tabular} & 8975044            & 5086959                                                           & \begin{tabular}[c]{@{}l@{}}4: 2543464\\ 5: 13\end{tabular}             & 0                                                                  \\ \hline
\begin{tabular}[c]{@{}l@{}}index36\_43\_1415\_\\ 1615.pcapng\end{tabular} & 8928621            & 4999994                                                           & \begin{tabular}[c]{@{}l@{}}4: 2499991\\ 5: 5\end{tabular}              & 0                                                                  \\ \hline
\begin{tabular}[c]{@{}l@{}}index44\_51.\\ pcapng\end{tabular}             & 9026447            & 5092084                                                           & \begin{tabular}[c]{@{}l@{}}4: 2546043\\ 5: 2\end{tabular}              & 0                                                                  \\ \hline
\begin{tabular}[c]{@{}l@{}}index\_52to59.\\ pcapng\end{tabular}           & 8982288            & 5180540                                                           & \begin{tabular}[c]{@{}l@{}}4: 2590267\\ 5: 5\end{tabular}              & 0                                                                  \\ \hline
\end{tabular}
}
\end{table}

    \item \textbf{[D3] \emph{Normal Dataset}}: This dataset comprises 13 network traffic capture files (1.7 GB) collected during normal operational runs under two distinct load conditions: 2kW and 16kW. The dataset includes MMS traffic between SCADA and PLC, as well as between SCADA and IED.
    \item \textbf{[D4] \emph{SIED\_pcaps} Dataset}: This dataset consists of network traffic packets captured during a session where a laptop in the SCADA network uses the IEC61850bean tool to query an Intelligent Electronic Device of Smart Home Zone of EPIC testbed.

  \item \textbf{[D5] \emph{state\_change\_bean\_and\_C\_code} Dataset}: This dataset has packet capture recorded on the EPIC testbed while an adversary exercised IEC 61850 control commands to operate a circuit breaker. The traffic includes MMS messages created by the iec61850bean tool and by a custom C code using the libiec61850 library. It captures the sequence of malicious control operations (open/close) and the device responses.


    \item \textbf{[D6] \emph{state\_change\_iec61850bean} Dataset}: This dataset includes MMS messages created by the iec61850bean tool to operate a circuit breaker. It captures the sequence of malicious control operations (open/close) and the device responses.

    \item \textbf{[D7] \emph{state\_change\_SCADA} Dataset}:
This dataset includes MMS messages sent to operate a circuit breaker using SCADA.
\end{itemize}

\section{Results}

\subsection{Analysis of MMS Read Operations in EPIC dataset}
The normal SCADA-to-IED and SCADA-to-PLC pairs represent standard authorized communication paths used for legitimate operations such as status monitoring, configuration checks, or retrieving measurement data. The \emph{domainId} and \emph{itemId} pairs found during normal operations are mentioned in Table~\ref{tab:epic_normalPairs}.
The domainIds like GIED1CTRL, TIED4MEAS (where G: Generation, T: Transmission, S: Smart Home and M: Microgrid stages) can be found in the request packets sent from SCADA to IED, followed by the itemIds LLN0\$DC\$NamPlt\$configRev, LLN0\$Measurement and LLN0\$Protection. The client periodically polls the LLN0\$DC\$NamPlt\$configRev attribute of each logical node, typically every 5 seconds, to detect configuration changes \cite{matouvsek2018description}. This attribute \emph{configRev} is a part of the LPL (Logical Node Nameplate), serves as a unique identifier of the logical device’s configuration. Any semantic change in the data model must update this value, ensuring that clients can reliably recognize when the interpretation of the data may have changed. LLN0\$Measurement and LLN0\$Protection correspond to measurement and protection datasets, respectively, which are also queried at regular intervals, where the protection dataset is queried to know the operation status (Op) of protection functions such as overvoltage and overcurrent which are under the logical nodes PTOV and PTOC. The measurement dataset is queried to know current magnitudes (phsA, phsB, phsC), total active power (TotW), reactive power (TotVAr), apparent power (TotVA), frequency (Hz), power factor (TotPF), phase-to-phase voltages (PPV phsAB, phsBC, phsCA), and phase-to-ground voltages (PhV phsA, phsB, phsC), which are under the Data Objects (DOs) of the Logical Node (LN) MMXU1. We could understand the data models of these datasets by inspecting the request response packets with MMS service code 12.
These are expected interactions between supervisory systems and IEDs in a secure, operational environment as shown in Table~\ref{tab:epic_normalPairs}. 

We also see MMS read operation packets from SCADA to PLCs of all stages in EPIC, which query the datasets corresponding to specific PLCs like AMI, CircuitBreaker, Sync, Testbed, VSD1 as shown in Table~\ref{tab:epic_normalPairs}, to read values of all the structures inside these datasets, which lie under LLN0 logical node to continuously know the device state during normal operating scenario.

In contrast, the pairs identified by the IEC61850bean tool include deeper or less typical accesses, such as accessing control functions (CF\$Mod), reporting behavior (RP\$urcbA01), or sensitive health and proxy data (ST\$PhyHealth, ST\$Proxy). These accesses may not align with standard SCADA behavior and can be abused to alter IED configurations, suppress alarms, reset counters, or flood networks, behaviors commonly associated with cyberattacks. Therefore, such IEC61850bean-to-IED access patterns are flagged as potential attack signatures and warrant closer scrutiny in a cybersecurity context.

Though, the IEC61850bean tool to IED pairs listed originate from the netwok traffic consisting of legitimate read operations collected as a part of loading data model of IEDs specific to the Smart Home zone of EPIC testbed. Such tools systematically browse and load the data model of the IED by reading all accessible nodes and attributes including control (CO,CF), status (ST), report (RP, EX), and configuration (DC) objects. So, while the access paths like CF\$Mod, RP\$urcbA01, or ST\$PhyHealth might look suspicious in a runtime operational context, they may in fact be benign during IED model exploration.

However, the reason they're flagged as "potential attack signatures" is because if similar access patterns occur unexpectedly during normal SCADA operations, especially outside commissioning or maintenance windows, they may indicate malicious activity such as unauthorized reads for reconnaissance or preparing for misconfiguration. So, it is not the access itself that is inherently malicious but it is the context (when, how often, and by whom it is accessed) that determines risk.

Hence, IEC61850bean tool generated pairs can be found both in read and write operations in the dataset under consideration, they can signal suspicious behavior and are therefore need to be monitored as they can be potential indicators of compromise.

\subsection{Analysis of MMS Write Operations in EPIC dataset}

Table~\ref{tab:mgmt_whitelist} shows the domainId and ItemId found in the packets requesting write operation.

During normal interval, we observe write operation request packets from SCADA (172.18.5.60) to PLCs (172.16.x.41), rather than to protection IEDs. This aligns with common engineering practice: time-critical interlocks and control logic are implemented locally in PLCs, validating prerequisites before execution, while fast trips and protection actions remain entirely within the IEDs and GOOSE messaging. By writing to PLCs/BCUs, SCADA can safely supervise and issue commands without risking delays or inadvertent changes to protection settings~\cite{Sheikh2025_SCADAProtectionIntegration}.

\begin{figure}[ht]
    \centering
    \begin{minipage}[t]{0.47\textwidth}
        \centering
        \includegraphics[width=\textwidth]{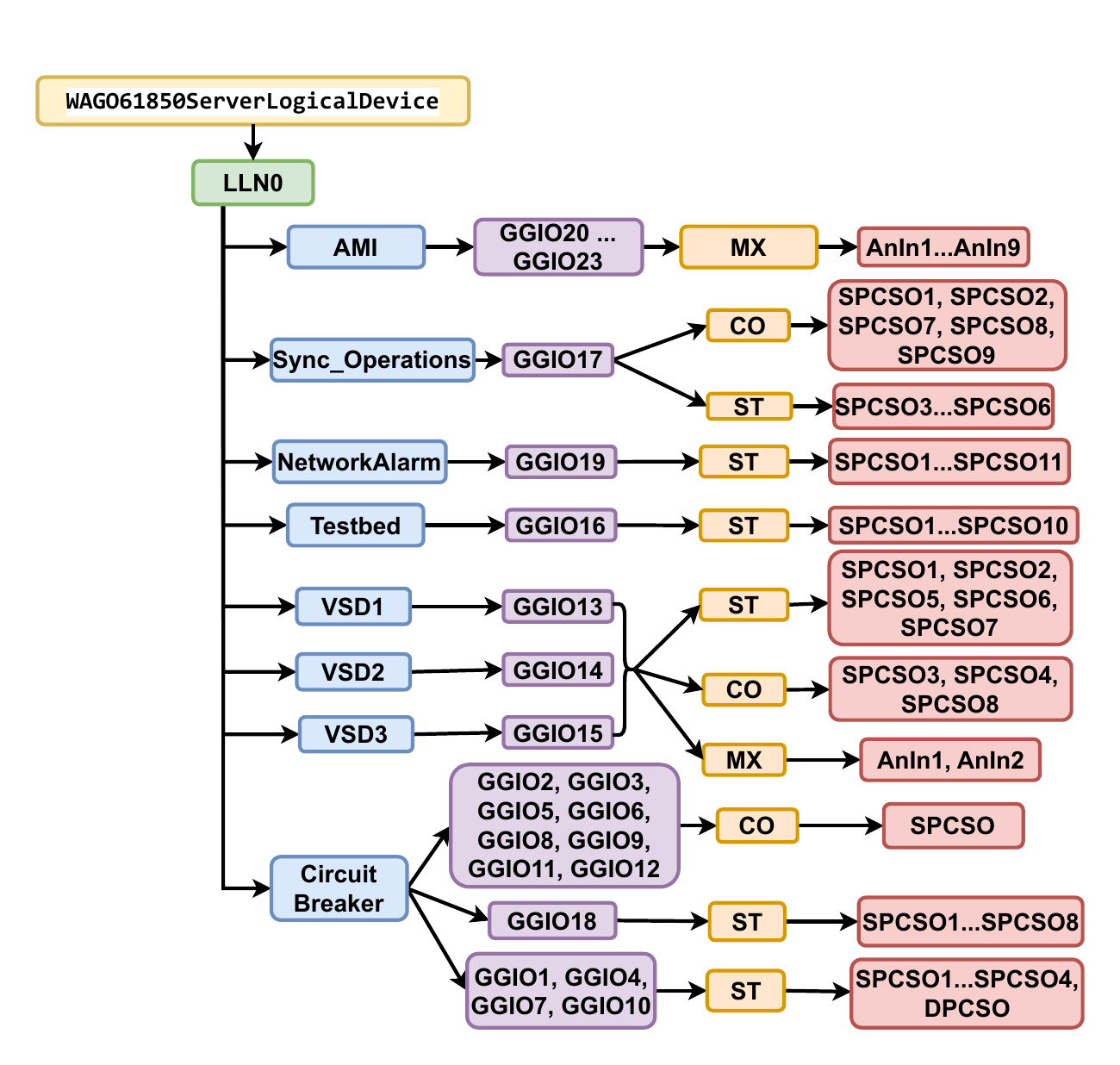}
        \caption{Mapping of GGIO Logical Nodes to Datasets, observed in GetDatasetDirectoryResponse packets.}
        \Description{Mapping of GGIO Logical Nodes to Datasets, observed in GetDatasetDirectoryResponse packets.}
        \label{fig:GGIO_dataset}
    \end{minipage}\hfill
    \begin{minipage}[t]{0.47\textwidth}
        \centering
        \includegraphics[width=\textwidth]{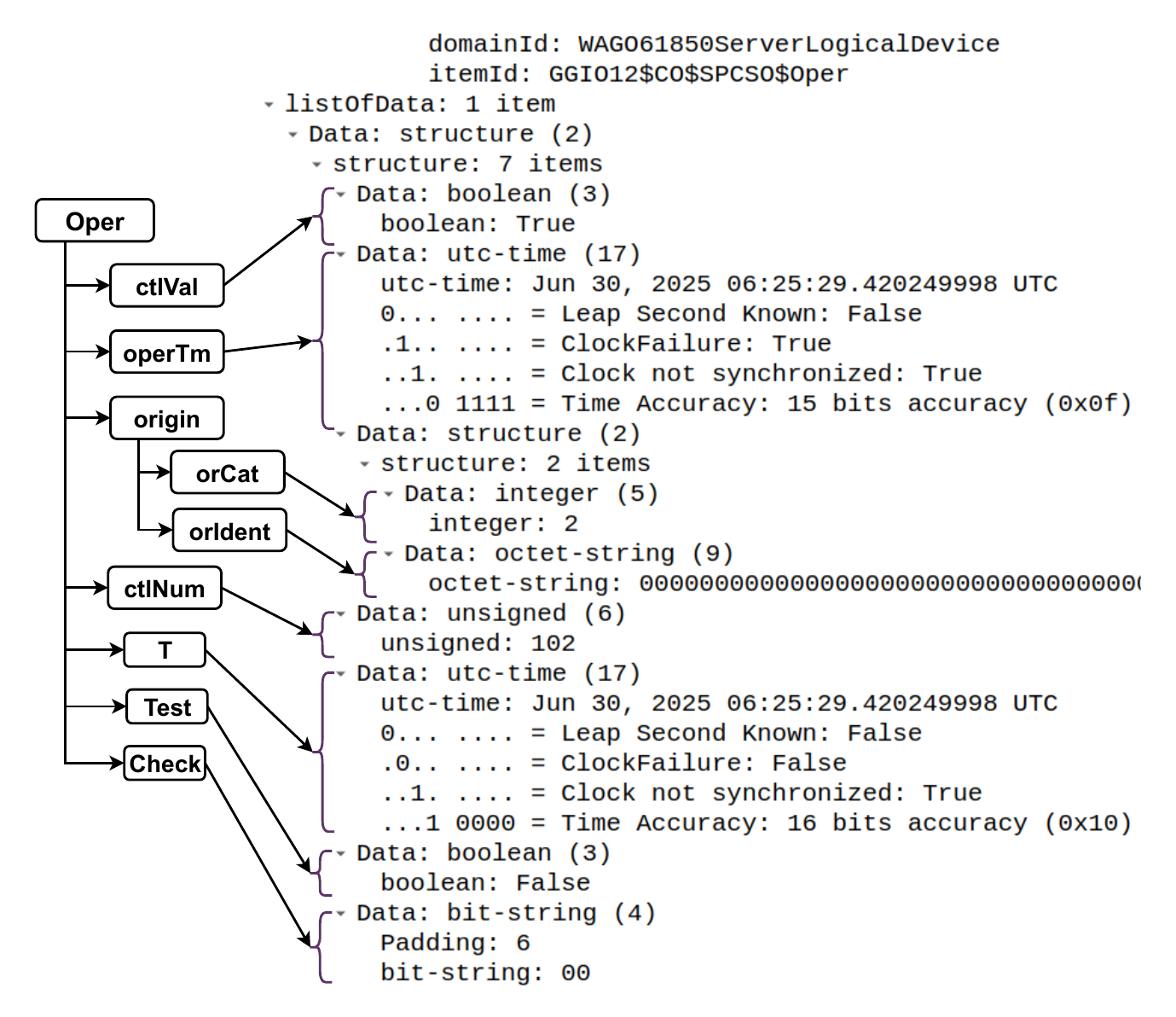}
        \caption{Oper structure mapping with the structure passed in write operation packet as seen in wireshark.}
        \Description{Oper structure mapping with the structure passed in write operation packet as seen in wireshark.}
        \label{fig:Oper_mapping}
    \end{minipage}
\end{figure}

We observe write operation request packets with domainId "WAGO61850ServerLogicalDevice" and itemId "GGIOx\$CO\$SPCSOy\$Oper". The itemId implies the operate command to a single-point control signal in a Generic I/O logical node GGIO logical node corresponds to generic process I/O. We barely understand which physical process is being written. Upon inspecting the GetDatasetDirectoryRequest packets and corresponding response packets, we could map these generic logical nodes to their parent datasets as shown in Figure~\ref{fig:GGIO_dataset}. This aids in understanding which physical process is the target of write operation.


The itemIds from IP addresses other than SCADA also look similar to the valid itemIds observed in the write operations from SCADA. We cannot clearly identify the attack packets carrying out write operations from that of the write operations carried out by the operator over SCADA. Hence we further inspect the payload of write request packet. As we understand from Table~\ref{tab:mgmt_whitelist} that all the write operations have similar itemId i.e. "GGIOx\$CO\$SPCSOy\$Oper", we inspect the Oper structure in depth in order to find exact attack signature.

IEC 61850 Control Model defines services like Operate (Oper) to change the state of physical processes by a client~\cite{code2010communication}. For example, to change the position of a switch on the peripheral device. 
The Operate service is utilized in conjunction with the control model, which defines mechanisms such as state machines, select-before-operate procedures, and time-activated control. Control service models rely on a specialized information model that employs controllable Common Data Classes (CDCs), for example, the SPC (Controllable Single Point). This model integrates attributes defined by the CDC alongside the service parameter ctlVal. Within the MMS protocol, these elements are represented as a structured variable, Oper, which includes the following components: ctlVal, operTm, origin.Orcat, origin.orIdent, ctlNum, T, Test, and Check as shown in Figure~\ref{fig:Oper_mapping}. At the MMS level, all these components must be written for execution of the Operate service, even in cases where only ctlVal is required. Nevertheless, all the fields (components) of Oper structure except “ctlVal” itself are filled by IEC-61850 driver automatically.

We observe that write operation packets from SCADA and attacker with same domain ID and itemID, for example, "WAGO61850\_ ServerLogicalDevice :: GGIO12\$CO\$SPCSO\$Oper", carry ListOfData in their payloads which consists of values of Oper structure and its components. We observe that \emph{Time Accuracy} under components \emph{operTm} and \emph{T} has values (0x0f,0x00/0x0f,0x11/0x0f,0x10) in the SCADA packet whereas the value in attacker packet sent using IEC61850bean tool is (0x0a,0x0a), and sent using libiec61850 attack script is (0x0a,0x00). Also the \emph{octet-string} under \emph{origin.orIdet} component is <MISSING> in case of attacker packet sent using libiec61850 attack script. The 0 bits accuracy and orIdent missing signify that the control command does not originate from a legitimate source and lacks proper attribution. This anomaly serves as a strong indicator of malicious activity, since valid control operations are always required to include a well-defined orIdent field to ensure traceability and accountability of the issuing entity. Also the \emph{origin.orCat} value 2 as in SCADA packet indicates station-control whereas value 3 as in attacker packet indicates remote-control~\cite{IEC61850}. Through which we can deduce that attack was launched remotely to send control command to a physical process.

Following fields correctly identify the attack packets in both the simulated traffic and real testbed traffic: 
\\
 \texttt{mms.confirmedServiceRequest == 5 \&\&   mms.iec61850.timeaccuracy == 0x00 \\ \&\& mms.data.octet-string=="" }



\subsection{Results M1: Attack Signature Extraction}
As per the steps mentioned in Section~\ref{M1}, we have considered [D1], [D2] [D3] and [D7] datasets having only normal traffic, as benign (seen) network trace. We have considered [D4], [D5] and [D6] having both normal and attack traffic as attack (unseen) network trace.

First we considered benign network traffic and extracted \emph{ip.src}, \emph{ip.dst}, \emph{mms.domainId}, \emph{mms.itemId}, \emph{mms.confirmedServiceRequest}, \emph{mms.iec61850.timeaccuracy} (this field retrieves two values for operTm and T components in Oper structure) and \emph{mms.data.octet-string} field values using tshark. We separated read and write operation packets into separate categories based on \emph{mms.confirmedServiceRequest} field values (read=4 and write=5). We obtained read whitelist, consisting of unique \emph{mms.domainId}, \emph{mms.itemId} pairs as shown in Table~\ref{tab:epic_normalPairs}, and write whitelist consisting of unique four tuples \emph{mms.domainId}, \emph{mms.itemId}, \emph{mms.iec61850.timeaccuracy}, and \emph{mms.data.octet-string} as shown in Table~\ref{tab:mgmt_whitelist}.

We repeated same steps for attack unseen network traffic, obtained potential attack read list and potential attack write list and compared them with respective read and write whitelist. We observed that the set of \emph{mms.domainId}, \emph{mms.itemId} pairs extracted from MMS read operations in normal traffic differs from those defined in the whitelist as shown in Table~\ref{tab:EPIC_analysis}. Only a limited subset of pairs, such as \emph{(SIEDyPROT, LLN0\$Measurement)} where, (y = 1, 2, 3, 4), overlap with the whitelist, indicating that these represent expected and legitimate read accesses to commonly polled logical nodes. Since IEC 61850-based SCADA systems typically perform read operations in a predictable and stable manner, deviations from the established set of valid \emph{domainId–itemId} pairs may indicate misconfiguration, abnormal behavior, or unauthorized access attempts. Therefore, we use the whitelist as a baseline to detect and raise alerts for MMS read packets whose \emph{domainId–itemId} pairs do not conform to the expected communication profile.

In case of write operation we could see the tuples that have different values against the field values dominantly in \emph{mms.iec61850.timeaccuracy} and \emph{mms.data.octet-string} among the four tuples. that are different from the ones found in normal SCADA write operations as shown in Table~\ref{tab:attack_sig}.
After inspecting the values found against the tuples in attack write operations we extracted following field value pairs that indicate the precise attack signatures for identifying the packets launched using IEC61850bean tool and libiec61850 based script. 
\begin{itemize}
\item \textbf{Attack Signature for IEC61850bean tool}:\\
 \texttt{mms.confirmedServiceRequest == 5 \&\&   mms.iec61850.timeaccuracy == 0x0a, 0x0a \\ \&\& mms.data.octet-string=="000...(64B)" } 
\item \textbf{Attack Signature for libiec61850 based attack script}:\\
 \texttt{mms.confirmedServiceRequest == 5 \&\&   mms.iec61850.timeaccuracy == 0x0a, 0x00 \\ \&\& mms.data.octet-string=="" } 
\end{itemize}

\begin{table}[H]
\centering
\caption{\emph{domainId} and \emph{itemId} pairs found during normal operating scenario in read packets.(Read Whitelist)}
\label{tab:epic_normalPairs}
\scalebox{0.9}{
\begin{tabular}{|l|l|}
\hline
\textbf{domainId (LD)} &
\textbf{itemid (LN\$FC\$DO\$DA)} \\ \hline
\begin{tabular}[c]{@{}l@{}}GIED(1/2)(CTRL/MEAS/DR)\\ TIED(1/2/4)(CTRL/MEAS/DR) \\ MIED(1/2)(CTRL/MEAS/DR)\\ SIED(1/2/3/4)(CTRL/MEAS/DR)\end{tabular} &
LLN0\$DC\$NamPlt\$configRev \\ \hline
\begin{tabular}[c]{@{}l@{}}GIED(1/2)PROT\\ TIED(1/2/4)PROT\\ MIED(1/2)PROT\\ SIED(1/2/3/4)PROT\end{tabular} &
\begin{tabular}[c]{@{}l@{}}LLN0\$DC\$NamPlt\$configRev\\ LLN0\$Measurement\\ LLN0\$Protection\end{tabular} \\ \hline
WAGO61850ServerLogicalDevice &
\begin{tabular}[c]{@{}l@{}}LLN0\$AMI\\ LLN0\$AT\_Signal\\ LLN0\$CircuitBreaker\\ LLN0\$DC\$NamPlt\$configRev\\ LLN0\$Gen\_Control\\ LLN0\$Loadbank1\_IN\\ LLN0\$Loadbank1\_OUT\\ LLN0\$Loadbank2\_IN\\ LLN0\$Loadbank2\_OUT\\ LLN0\$Microgrid\_Control\\ LLN0\$Network\\ LLN0\$NetworkAlarm\\ LLN0\$Q1ASync\\ LLN0\$SwitchAlarm\\ LLN0\$Sync\\ LLN0\$Sync\_Operations\\ LLN0\$Testbed\\ LLN0\$VSD1\\ LLN0\$VSD2\\ LLN0\$VSD3\end{tabular} \\ \hline
\end{tabular}
}
\end{table}

\begin{table}[H]
\centering
\caption{Whitelist of write operations. (Write Whitelist)}
\label{tab:mgmt_whitelist}
\scalebox{0.9}{
\begin{tabular}{|l|l|l|l|}
\hline
\textbf{DomainId} &
  \textbf{ItemId} &
  \textbf{\begin{tabular}[c]{@{}l@{}}Time \\ Acc.\end{tabular}} &
  \textbf{\begin{tabular}[c]{@{}l@{}}origin.\\ OrIdent\\ (64 Bytes)\end{tabular}} \\ \hline
\multirow{3}{*}{\begin{tabular}[c]{@{}l@{}}WAGO61850Server\\ LogicalDevice\end{tabular}} &
  \begin{tabular}[c]{@{}l@{}}GGIO17\$CO\$SPCSO1\$Oper\\ GGIO17\$CO\$SPCSO2\$Oper\\ GGIO5\$CO\$SPCSO\$Oper\\ GGIO13\$CO\$SPCSO1\$Oper\end{tabular} &
  \begin{tabular}[c]{@{}l@{}}0x0f,\\ 0x00\end{tabular} &
  \multirow{3}{*}{000...} \\ \cline{2-3}
 &
  GGIO15\$CO\$SPCSO2\$Oper &
  \begin{tabular}[c]{@{}l@{}}0x0f,\\ 0x11\end{tabular} &
   \\ \cline{2-3}
 &
  \begin{tabular}[c]{@{}l@{}}GGIO4\$CO\$SPCSO4\$Oper\\ GGIO2\$CO\$SPCSO\$Oper\\ GGIO17\$CO\$SPCSO8\$Oper\\ GGIO5\$CO\$SPCSO\$Oper\\ GGIO13\$CO\$SPCSO4\$Oper\\ GGIO4\$CO\$SPCSO6\$Oper\\ GGIO2\$CO\$SPCSO3\$Oper\\ GGIO15\$CO\$SPCSO3\$Oper\\ GGIO13\$CO\$SPCSO3\$Oper\\ GGIO3\$CO\$SPCSO\$Oper\\ GGIO9\$CO\$SPCSO\$Oper\\ GGIO17\$CO\$SPCSO2\$Oper\\ GGIO6\$CO\$SPCSO\$Oper\\ GGIO12\$CO\$SPCSO\$Oper\\ GGIO15\$CO\$SPCSO1\$Oper\\ GGIO8\$CO\$SPCSO\$Oper\\ GGIO13\$CO\$SPCSO\$Oper\\ GGIO4\$CO\$SPCSO5\$Oper\\ GGIO2\$CO\$SPCSO6\$Oper\\ GGIO2\$CO\$SPCSO5\$Oper\\ GGIO17\$CO\$SPCSO1\$Oper\\ GGIO14\$CO\$SPCSO3\$Oper\\ GGIO2\$CO\$SPCSO4\$Oper\\ GGIO11\$CO\$SPCSO\$Oper\\ GGIO4\$CO\$SPCSO7\$Oper\\ GGIO4\$CO\$SPCSO3\$Oper\\ GGIO13\$CO\$SPCSO1\$Oper\\ GGIO17\$CO\$SPCSO7\$Oper\end{tabular} &
  \begin{tabular}[c]{@{}l@{}}0x0f,\\ 0x10\end{tabular} &
   \\ \hline
\end{tabular}
}
\end{table}

\begin{table}[H]
\centering
\caption{Results of comparison of [D4] and [D6] attack network traffic with Read Whitelist. (Only few are listed).}
\label{tab:EPIC_analysis}
\scalebox{0.9}{
\begin{tabular}{|c|l|}
\hline
\textbf{Domain Id (LD)} & \textbf{Item Id (LN\$FC\$DO\$DA)} \\ \hline
\begin{tabular}[c]{@{}c@{}}SIEDyZ(y=1/2/3/4),\\ Z=CTRL/DR/MEAS/PROT\end{tabular} &
\begin{tabular}[c]{@{}l@{}}%
BI6GGIO1\$CF\$Mod \\ 
BI6GGIO1\$DC\$NamPlt \\ 
\textcolor{blue}{LLN0\$Measurement} \\ 
LPHD1\$RP\$urcbA01 \\ 
LPHD1\$RP\$urcbB01 \\ 
LPHD1\$ST\$PhyHealth \\ 
LPHD1\$ST\$Proxy \\ 
PTRC1\$CF\$Mod \\ 
PTRC1\$CF\$OpCntRs \\ 
PTRC1\$DC\$NamPlt \\ 
PTRC1\$EX\$urcbA01
\end{tabular} \\ \hline
\end{tabular}
}
\end{table}

\begin{table}[H]
\caption{Results of comparison of [D5] and [D6] with write whitelist}
\label{tab:attack_sig}
\scalebox{0.9}{
\begin{tabular}{|l|l|l|l|l|l|}
\hline
\textbf{Dataset}          & \textbf{sip-\textgreater{}dip}                                                     & \textbf{DomainId}                                                          & \textbf{ItemId}                                                          & \textbf{\begin{tabular}[c]{@{}l@{}}Time\\ Acc.\end{tabular}} & \textbf{\begin{tabular}[c]{@{}l@{}}origin.\\ orIdent\end{tabular}} \\ \hline
\multirow{3}{*}{{[}D5{]}} & \begin{tabular}[c]{@{}l@{}}172.16.5.103\\ -\textgreater\\ 172.16.3.41\end{tabular}  & \begin{tabular}[c]{@{}l@{}}WAGO61850\\ ServerLogical\\ Device\end{tabular} & \begin{tabular}[c]{@{}l@{}}GIO12\$CO\$\\ SPCSO\$Oper\end{tabular}          & \begin{tabular}[c]{@{}l@{}}0x0a,\\ 0x00\end{tabular}         & \textless{}MISSING\textgreater{}                                   \\ \cline{2-6} 
                          & \begin{tabular}[c]{@{}l@{}}172.16.4.201\\ -\textgreater\\ 172.16.3.41\end{tabular} & \begin{tabular}[c]{@{}l@{}}WAGO61850\\ ServerLogical\\ Device\end{tabular} & \begin{tabular}[c]{@{}l@{}}GIO12\$CO\$\\ SPCSO\$Oper\end{tabular}          & \begin{tabular}[c]{@{}l@{}}0x0a,\\ 0x0a\end{tabular}         & 000...                                                             \\ \cline{2-6} 
                          & \begin{tabular}[c]{@{}l@{}}172.16.4.201\\ -\textgreater\\ 172.16.3.41\end{tabular} & \begin{tabular}[c]{@{}l@{}}WAGO61850\\ ServerLogical\\ Device\end{tabular} & \begin{tabular}[c]{@{}l@{}}GIO12\$CO\$\\ SPCSO\end{tabular}                & \begin{tabular}[c]{@{}l@{}}0x0a,\\ 0x0a\end{tabular}         & 000...                                                             \\ \hline
\multirow{2}{*}{{[}D6{]}} & \begin{tabular}[c]{@{}l@{}}172.16.4.201\\ -\textgreater\\ 172.16.3.41\end{tabular} & \begin{tabular}[c]{@{}l@{}}WAGO61850\\ ServerLogical\\ Device\end{tabular} & \begin{tabular}[c]{@{}l@{}}GIO12\$CO\$\\ SPCSO\$Oper\\ \$ctlVal\end{tabular} & \begin{tabular}[c]{@{}l@{}}0x0a,\\ 0x0a\end{tabular}         & 000...                                                             \\ \cline{2-6} 
                          & \begin{tabular}[c]{@{}l@{}}172.16.4.201\\ -\textgreater\\ 172.16.3.41\end{tabular} & \begin{tabular}[c]{@{}l@{}}WAGO61850\\ ServerLogical\\ Device\end{tabular} & \begin{tabular}[c]{@{}l@{}}GIO12\$CO\$\\ SPCSO\$Oper\end{tabular}          & \begin{tabular}[c]{@{}l@{}}0x0a,\\ 0x0a\end{tabular}         & 000...                                                             \\ \hline
\end{tabular}
}
\end{table}

\subsection{Results M2: Attack and Attack Path Detection}
We incorporated the identified attack signatures for write operations and read whitelist in our IDS and inspected across all the pcaps in attack datasets [D5], and [D6]. We found that our module detects all the attacks instances and optputs the attacks paths as shown in Table~\ref{tab:attack_sig}. The attacks are launched by IEC61850bean tool (172.16.5.103) and libiec61850 attack script (172.16.5.103) have targeted Smart home zone PLC to change the state of circuit breaker device as reflected in the itemID field. As per Figure~\ref{fig:GGIO_dataset} GGIO12 logical node belongs to the circuit breaker dataset.  

\section{Conclusions}
This work demonstrates that MMS-based communication in IEC 61850 smart substations, while essential for efficient SCADA operations, introduces critical cyber-attack surfaces that can be exploited to manipulate process states and disrupt grid reliability. By performing an in-depth analysis of the MMS protocol and systematically extracting security-relevant field–value pairs, the proposed framework enables fully automated detection and prevention of remote cyberattacks targeting PLCs and IEDs. Validation on seven datasets—covering normal operation as well as IEC61850Bean-based and libiec61850 script-driven attacks—shows that the framework accurately identifies legitimate-looking yet malicious MMS packets, particularly those aimed at unauthorized circuit breaker manipulation in the EPIC testbed. The results confirm that the proposed approach effectively distinguishes benign from attack-signature-carrying MMS traffic, thereby significantly improving situational awareness and cyber resilience of IEC 61850-compliant smart substations against sophisticated remote attacks.
\begin{acks}

\end{acks}

\bibliographystyle{ACM-Reference-Format}
\bibliography{sample-base}

\appendix

\end{document}